\documentclass[useAMS,usenatbib]{mn2e}

\usepackage{graphicx}
\usepackage{amsmath}
\usepackage{amsfonts}
\usepackage{amssymb}
\usepackage{rotating}
\usepackage[english]{babel}

\title[SPH  simulations  of  the   core-degenerate  scenario  for  SNe
  Ia]{Smoothed    Particle    Hydrodynamics   simulations    of    the
  core-degenerate scenario for Type Ia supernovae}

\author[G. Aznar--Sigu\'an et al.]{G. Aznar--Sigu\'an$^{1,2}$,
                                E. Garc\'\i a--Berro$^{1,2}$\thanks{Corresponding author, email: enrique.garcia-berro@upc.edu},
                                P. Lor\'en--Aguilar$^{3}$
				N. Soker$^{4}$,
                                A. Kashi$^5$\\
       $^1$Departament de F\'\i sica Aplicada, 
           Universitat Polit\`ecnica de Catalunya,
           c/Esteve Terrades 5, 
           08860 Castelldefels, 
           Spain\\
       $^2$Institute for Space  Studies of Catalonia,
           c/Gran Capit\`a 2--4, Edif. Nexus 201,   
           08034  Barcelona, 
           Spain\\
       $^3$School of Physics, 
           University of Exeter, 
           Stocker Road, 
           Exeter EX4 4QL, UK\\
       $^4$Department of Physics, 
           Technion – Israel Institute of Technology, 
           Haifa 32000, 
           Israel\\
       $^5$Minnesota Institute for Astrophysics, 
           University of Minnesota,
           116 Church St. SE., 
           Minneapolis, MN 55455, 
           USA}

\begin{document}

\date{\today}

\maketitle

\begin{abstract}
The  core-degenerate (CD)  scenario  for type  Ia  supernovae (SN  Ia)
involves the  merger of  the hot  core of  an asymptotic  giant branch
(AGB) star  and a white  dwarf, and might contribute  a non-negligible
fraction  of  all  thermonuclear supernovae.   Despite  its  potential
interest, very few  studies, and based on  only crude simplifications,
have been devoted to investigate this possible scenario, compared with
the large  efforts invested  to study some  other scenarios.   Here we
perform the  first three-dimensional simulations of  the merger phase,
and find  that this  process can  lead to the  formation of  a massive
white dwarf, as required by this scenario. We consider two situations,
according  to the  mass of  the  circumbinary disk  formed around  the
system during the  final stages of the common envelope  phase.  If the
disk is massive  enough, the stars merge on a  highly eccentric orbit.
Otherwise,  the merger  occurs after  the circumbinary  disk has  been
ejected and gravitational  wave radiation has brought  the stars close
enough  for the  secondary  to  overflow its  Roche  lobe radius.  Not
surprisingly, the  overall characteristics of the  merger remnants are
similar  to  those  found  for the  double-degenerate  (DD)  scenario,
independently of the very different core temperature and of the orbits
of the merging  stars. They consist of a central  massive white dwarf,
surrounded  by a  hot,  rapidly  rotating corona  and  a thick  debris
region.
\end{abstract}



\section{Introduction}

Thermonuclear,  or Type  Ia supernovae  (SNe Ia),  originate from  the
explosion  of  carbon-oxygen  white   dwarfs  ---  see,  for  instance
\cite{Hillebrandt2013}  and \cite{pilar}  for  recent  reviews on  the
explosion mechanisms  and the different evolutionary  channels --- and
are among the  most luminous events in the  universe.  Moreover, since
they  arise from  the detonation  of  white dwarfs  with a  relatively
narrow  range  of masses,  SNe  Ia  typically have  similar  intrinsic
luminosities  that   can  be   conveniently  standardized   using  the
properties of the  light curves, and consequently play  a primary role
in cosmology.  However,  the precise nature of  the progenitor systems
that give rise to SNe Ia remains  still to be unveiled, and is a topic
of active  research. The fact that  several peculiar SNe Ia  have been
discovered  so far  (e.g., \citealt{Bildstenetal2007,  Peretsetal2010,
Jordanetal2012,  Foleyetal2013}) suggests  that different  progenitors
and/or explosion mechanisms could trigger such outbursts, resulting in
normal and peculiar SNe Ia. For  instance, the variety of peculiar SNe
Ia and their range of properties, has led to explore sub-Chandrasekhar
mass  explosions  ---  see,   for  instance,  \cite{Wangetal2013}  and
\cite{Scalzoetal2014}.

It has long been  suggested that a white dwarf in  a binary system ---
with   either  another   white   dwarf  ---   through  the   so-called
double-degenerate (DD)  scenario \citep{Webbink1984, IbTu84} ---  or a
main-sequence or red giant companion --- through the single-degenerate
(SD) scenario  \citep{Whelan1973, Nomoto1982, Han2004} ---  could give
rise to a SN  Ia event.  More recent scenarios ---  see the reviews of
\cite{WangHan2012} and \cite{Maozetal2014} --- are variants of the two
previously mentioned ones, and include the double detonation scenario,
where  helium-rich material  is  accreted by  the carbon-oxygen  white
dwarf (e.g.,  \citealt{ShenBildsten2009, Ruiteretal2011}),  the double
white dwarf collision scenario \citep{Raskinetal2009, Rosswogetal2009,
Thompson2011, Kushniretal2013}, and  the core-degenerate (CD) scenario
\citep{Sparks1974,   Livio2003,   Kashi2011,   Ilkov2012,   Ilkov2013,
Sokeretal2013} to be  studied here.  Within the CD  scenario the white
dwarf merges  with the hot core  of a massive asymptotic  giant branch
(AGB) star  \citep{Livio2003, Kashi2011, Ilkov2012, Ilkov2013}.   In a
recent  paper  \cite{TsebrenkoSoker2015a}   argue  that  approximately
$20\%$, and likely many more, of all SNe Ia come from the CD scenario.
However, \cite{2015arXiv150203898Z} and  \cite{Meng} using Monte Carlo
binary population synthesis codes concluded  that the fraction is only
2--10\%, but  another recent population synthesis  study suggests that
the merger of the core of an AGB  star and a white dwarf can indeed be
very  common   \citep{Briggsetal2015}.   A  table   summarizing  these
scenarios with  their main  advantages and drawbacks  can be  found in
\cite{TsebrenkoSoker2015a}. To  these we  add a  note on  the recently
proposed singly-evolved star (SES) scenario \citep{Chiosi}.

The DD and the CD scenarios share many similarities, and in particular
they have a common initial evolutionary path.  Typically, in a low- or
intermediate-mass binary  system the more massive  (primary) component
fills  its  Roche lobe  while  crossing  the  Hertzsprung gap  in  the
color-magnitude diagram and  the mass transfer episode  is stable. The
primary star  then evolves into  a carbon-oxygen white  dwarf.  Later,
the initially secondary  star fills its Roche lobe when  it becomes an
evolved giant  star.  This time  the combination of a  deep convective
envelope and  an extreme mass  ratio causes an unstable  mass transfer
episode from  the giant to  the white dwarf  and a common  envelope is
formed.  In the DD channel the common envelope is entirely ejected and
a final binary system made of  two white dwarfs is obtained, where the
second  white   dwarf  formed  is   normally  the  more   massive  one
\citep{Webbink1984}.  Gravitational wave radiation finally brings both
white dwarfs closer and a merger  occurs. In the CD channel the merger
occurs shortly after the common envelope phase.

Note, however, that there are  two significant differences between the
DD and the CD scenarios.  In a DD merger both white dwarfs are brought
together by gravitational wave radiation,  a process which lasts for a
long time.  Consequently,  it is expected that both  components of the
binary system are cool, and have nearly circular orbits.  Moreover, it
is  also foreseen  that both  white dwarfs  will be  synchronized.  In
contrast, in the CD scenario the  merger of both stars is triggered by
the interaction with the circumbinary  disk.  In particular, in the CD
scenario during the  final stages of the common  envelope phase, $\sim
10  \%$ of  the common  envelope remains  bound to  the binary  system
formed  by the  AGB  star and  the companion  white  dwarf, forming  a
circumbinary disk  --- see  \cite{Kashi2011}, and  references therein.
Consequently, it  turns out that in  a sizable number of  cases as the
binary system transfers angular momentum to the circumbinary disk, the
separation of  the pair decreases  and the eccentricity of  the system
increases \citep{Artymowicz1991}  while the  core of  the AGB  star is
still  hot \citep{Kashi2011},  favoring a  merger  at the  end of  the
common  envelope phase  or a  short time  after, during  the planetary
nebula  phase.  However,  it  might  be also  possible  that,  if  the
circumbinary disk is not massive  enough, the merger takes place after
the  circumbinary disk  has been  already ejected.   In this  case the
driver of the merger is again  the emission of gravitational waves, so
the orbits will  have a small eccentricity and the  temperature of the
core of the post-AGB star will be low.  In summary, in the CD scenario
a considerably large range of  eccentricities of the binary system and
of temperatures  of the naked  core of the  AGB star are  expected, at
odds with what it is foreseen for the DD channel.

While  several simulations  of  the DD  progenitor  channel have  been
already  performed   over  the   last  two   decades  \citep{Benz1990,
Segretain1997,   Gea04,   Yoon2007,   Loren-Aguilar2009,   Pakmor2010,
Dan2011, Raskin2013},  no full three-dimensional simulation  of the CD
scenario has been done so far.  The present paper aims at filling this
gap,  and  it is  organized  as  follows. In  Section~\ref{sec:ic}  we
discuss the initial conditions considered to simulate the CD scenario.
It  is  followed  by  Section~\ref{sec:numerical},  where  we  briefly
explain  the numerical  techniques  used in  this  paper.  In  Section
\ref{sec:results} we describe in depth the results of our simulations.
Finally,  in   Section~\ref{sec:conclusion}  we  summarize   our  main
findings, and we draw our conclusions.

\section{Initial conditions}
\label{sec:ic}

\begin{figure*}
   \begin{center}
      {\includegraphics[width=0.47\textwidth, trim=0.1cm 0.1cm 0.1cm 0.1cm,clip=true]{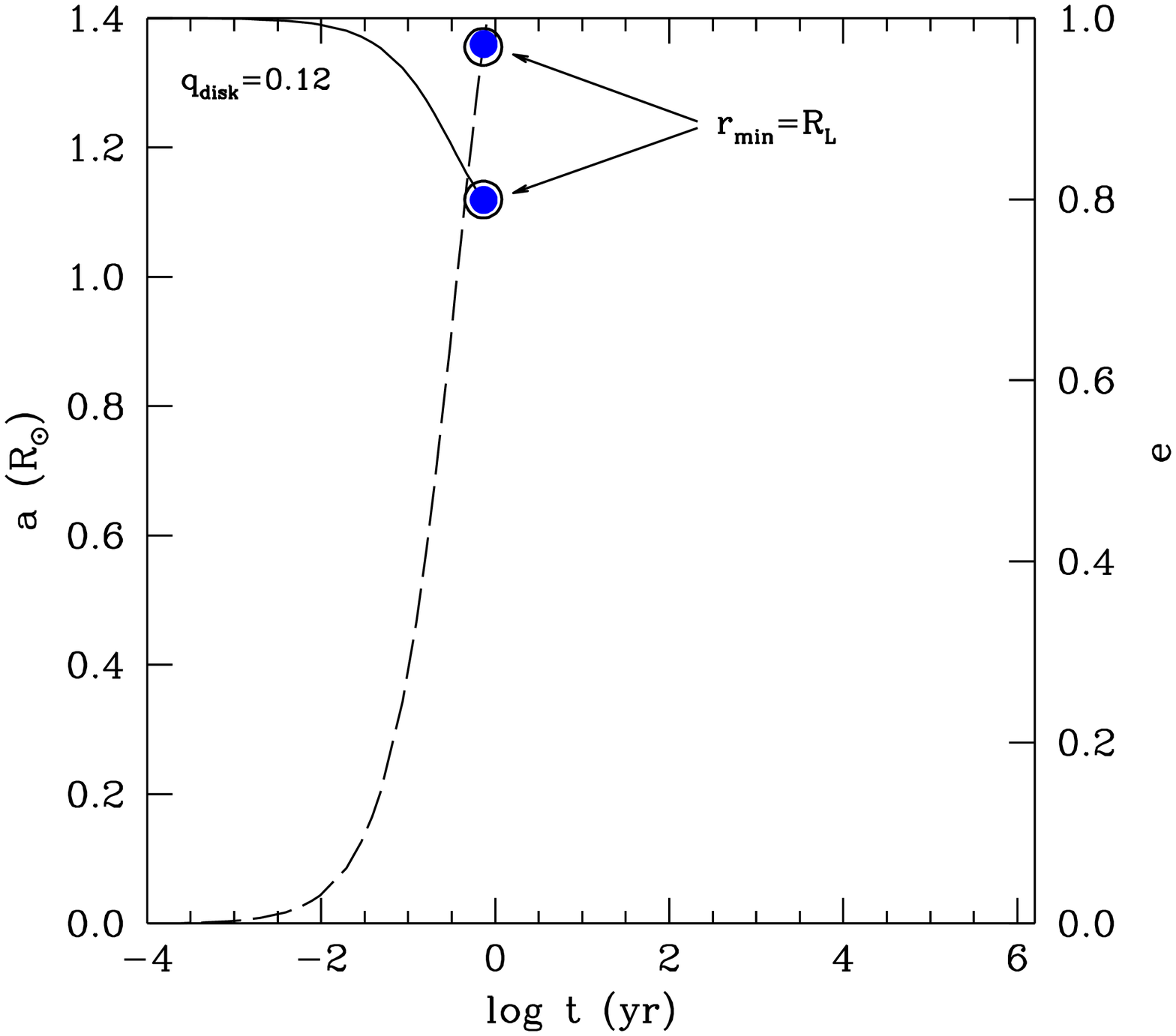}\hspace{0.4cm}
       \includegraphics[width=0.47\textwidth, trim=0.1cm 0.1cm 0.1cm 0.1cm,clip=true]{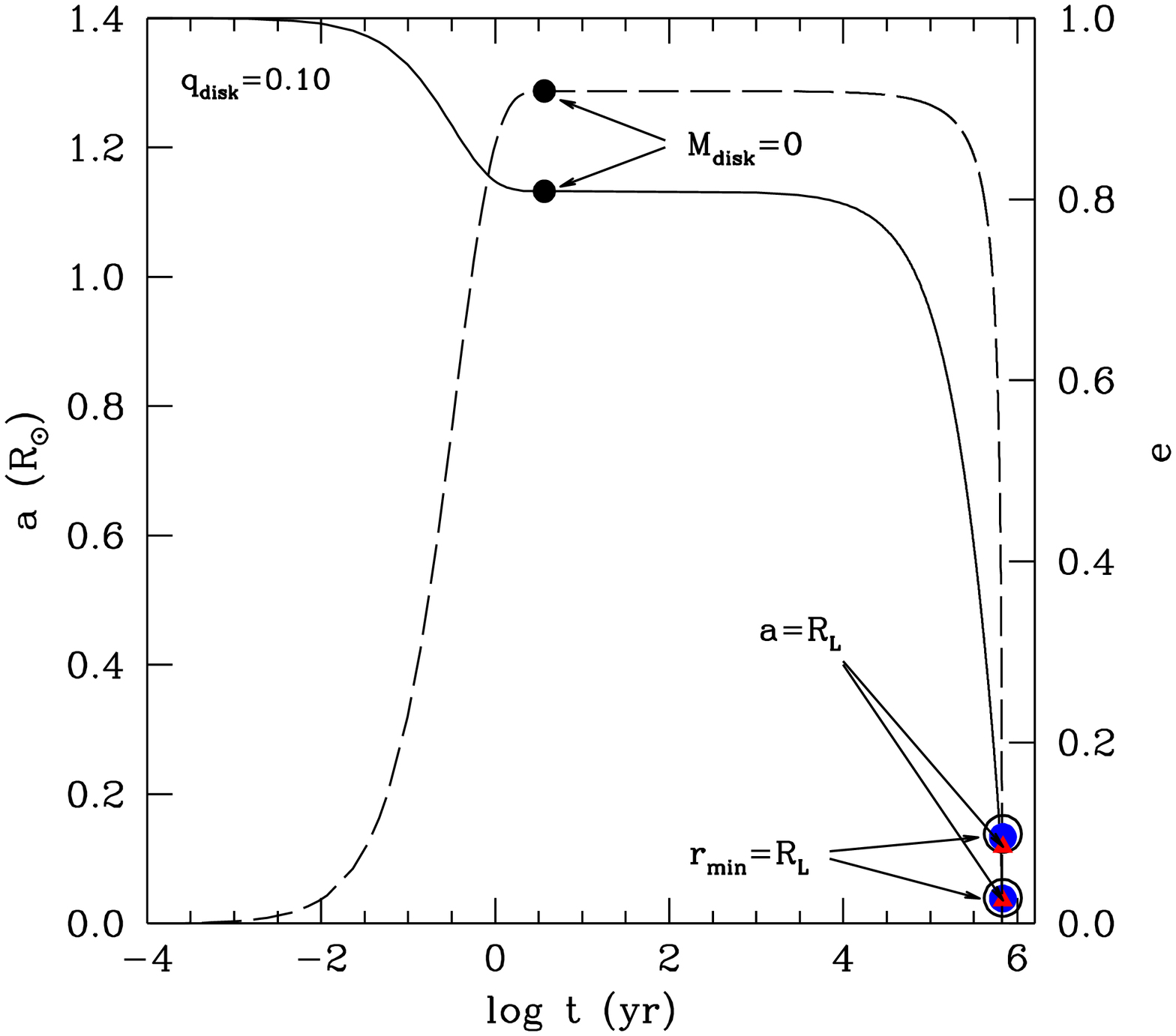}}\\
    \caption{Evolution of the orbital  parameters of the binary system
      formed by an  AGB star and a white dwarf  during its interaction
      with  the  circumbinary  disk,  once a  large  fraction  of  the
      envelope of the AGB star has  been ejected. The left panel shows
      the  case in  which  a massive  circumbinary  disk with  $q_{\rm
      disk}=0.12$  is adopted,  whereas the  right panel  the case  in
      which  $q_{\rm  disk}=0.10$  is  considered  ---  see  text  for
      details.  The solid  lines and left scale show  the evolution of
      the semimajor  axis ($a$),  and the dashed  lines and  the right
      scale display that of the eccentricity ($e$).  The hollow, large
      circles indicate the point at which the minimum orbital distance
      is such that the secondary would overflow its Roche lobe for the
      equivalent  circular  orbit, $r_{\rm  min}=a(1-e)=a_{\rm  ECO}$,
      whereas the  blue solid  symbols show  the orbital  distance $a$
      when the  secondary white dwarf  fills its Roche-lobe  radius in
      our simulations,  $r=r_{\rm O}$.  Finally, for  the right panel,
      the  point at  which the  semi-major axis  of the  binary system
      equals  the   distance  at  which  the   secondary  reaches  the
      Roche-lobe radius  for the equivalent circular  orbit, $a=a_{\rm
      ECO}$ is indicated  by a triangle.  Also in this  panel the time
      at  which the  circumbinary  disk has  been  totally ejected  is
      indicated by the solid black circles.}
   \label{fig:ic} 
   \end{center}
\end{figure*}

\subsection{The final phase of the common envelope}
\label{CE}

\cite{Kashi2011} estimated that  during the last stages  of the common
envelope phase, between $\sim 1\%$ and $10 \%$ of the ejected envelope
remains bound  to the binary system  composed of the cold  white dwarf
and the hot core of the AGB star. They furthermore suggested that, due
to  angular  momentum  conservation  and further  interaction  of  the
remaining  common  envelope  with  the  binary  system,  the  fallback
material  forms a  circumbinary disk  around the  binary.  This  newly
formed disk would later interact with the binary system, thus reducing
the orbital separation of the pair  much more than what it was reduced
during the dynamical phase where the common envelope is ejected, while
the  eccentricity of  the  binary system  would  increase.  Using  the
results of  \cite{Artymowicz1991}, \cite{Kashi2011} obtained  the rate
at  which the  semimajor  axis decreases  and the  rate  at which  the
eccentricity of the orbit increases.  In addition, they also took into
account the rate at which the circumbinary disk losses mass during the
interaction.

\cite{Kashi2011} found that the final values of the orbital parameters
critically  depend on  the amount  of matter  ejected from  the system
through  disk winds.   Specifically, they  found that  for disks  with
masses  $M_{\rm disk}  \ga  0.12(M_{\rm core}+M_{\rm  WD})$ ---  where
$M_{\rm disk}$,  $M_{\rm core}$,  and $M_{\rm WD}$  are, respectively,
the masses of the  disk, the core of the AGB star  and the white dwarf
--- the circumbinary disk-binary interaction leads  to a merger if the
speed of  the disk wind  is the  escape velocity.  For  the particular
case  studied by  \cite{Kashi2011}, where  $M_{\rm WD}=0.6\,M_{\sun}$,
$M_{\rm  core}=0.77\,M_{\sun}$ and  $M_{\rm env}=4.23\,M_{\sun}$,  the
mass  of  the  disk  amounts  to   $\sim  4\%$  of  the  mass  of  the
envelope. They also found that for less massive circumbinary disks the
separation is  not reduced  enough to  result in  a merger  during the
interaction. In these cases,  gravitational wave radiation will reduce
the   orbital   separation,   $a$,    and   the   eccentricity,   $e$,
further. Consequently, the final merger  would occur when the orbit of
the binary system is almost circular.   Hence, in these cases it turns
out  that the  orbital separation,  $a$, and  the distance  at closest
approach,  $r_{\rm min}=a(1-e)$,  are  very similar.   The first  mass
transfer episode occurs when $r_{\rm min}  < r_{\rm O}$ but $a> r_{\rm
O}$, being  $r_{\rm O}$ the  distance between  the two stars  when the
secondary   overflows  its   Roche  lobe   radius  in   our  numerical
simulations. It  is convenient to  define here an  equivalent circular
orbit, which  is the circular  orbit for which $R_2\simeq  R_{\rm L}$,
where $R_2$  is the radius of  the secondary star, and  $R_{\rm L}$ is
computed  using  the expression  of  \cite{Eggleton83}.   We call  the
semi-major  axis of  this orbit  $a_{\rm ECO}$.   Moreover, since  the
timescale for gravitational wave emission is long, the core of the AGB
star would  be cold  \citep{Kashi2011}.  Note, however,  that although
the  procedure detailed  in  \cite{Kashi2011} takes  into account  the
tidal interaction  during the  common envelope  phase, to  compute the
orbital parameters  of our system  at which our SPH  simulations start
(see  below) we  did  not take  into account  tidal  effects once  the
circumbinary disk has been totally ejected from the system.

\subsection{Orbital parameters}

We  compute the  evolution of  the  orbital parameters  of the  binary
system before  the merger  occurs following  closely the  procedure of
\cite{Kashi2011}.   In  particular, we  consider  the  same binary  as
\cite{Kashi2011} and we adopt two different disk masses.  Writing
\begin{equation}
M_{\rm  disk}=q_{\rm disk}(M_{\rm  core}+M_{\rm  WD}),
  \label{eq:qdisk}
\end{equation}
we adopt as  reference cases $q_{\rm disk}  = 0.12$ --- a  disk with a
mass just over  the critical one ---  and $q_{\rm disk} =  0.10$ --- a
low-mass one. When $q_{\rm disk} = 0.12$ the merger occurs just before
the disk is totally ejected from  the system.  Hence, the merger would
occur when the  orbit of the binary system is  still highly eccentric,
and moreover the core  of the AGB star would still  be hot, while when
$q_{\rm disk}=0.10$ the merger occurs long after the circumbinary disk
has been totally ejected from the  system, and hence the orbit will be
nearly circular, and the core of the AGB ill be cold.

The time evolution  of the orbital parameters  for $q_{\rm disk}=0.12$
is displayed in the left panel of Fig.~\ref{fig:ic}, where we show the
evolution of both the orbital separation (solid line, left scale), and
of the  eccentricity (dashed  line, right  scale).  The  large, hollow
circles  correspond  to  the  time instant  at  which  the  periastron
distance is  such that  the radius  of the  secondary would  equal the
Roche-lobe radius ($R_2=R_{\rm L}$) for the equivalent circular orbit,
while  the solid,  blue  circles  show the  point  at  which the  mass
transfer  episode begins  --- see  below  for details  about the  mass
transfer episode.  Note that both times are almost coincidential.  For
this disk,  the merger  occurs at  $t\sim 270$  days after  the common
envelope is ejected.

The equivalent diagram for the case  in which $q_{\rm disk} = 0.10$ is
shown in the right panel of  Fig.~\ref{fig:ic}. In this case, the disk
has been totally ejected at early times (the time at which this occurs
is  represented using  black, solid  circles in  this figure).   For a
relatively  long   time  interval  the  orbital   separation  and  the
eccentricity of the  pair remain almost constant, but as  time goes on
the  emission  of  gravitational  waves  progressively  decreases  the
orbital distance  and the eccentricity  of the pair, being  faster for
times  just before  the merger  begins.  Eventually,  at approximately
$6\times  10^{5}$ years  after the  common envelope  of the  system is
ejected both stars merge. As in the previous case we also show in this
panel the  points where  $R_2=R_{\rm L}$  for the  equivalent circular
orbit, and  the points at  which the merger starts.   Additionally, in
this panel  we also show using  red triangles the times  for which the
semi-major axis equals the distance at which the secondary reaches the
Roche  lobe  radius  for  the  equivalent  circular  orbit  ($a=a_{\rm
ECO}$). Note that in this case the eccentricity is small ($e\sim 0.1$)
and  that   we  expect  that,   given  the  very  long   timescale  of
gravitational wave emission, the post-AGB  star will have a cold core.
In cases where merger takes  place within several times $10^5~$yr, the
SN Ia  ejecta might catch  up with  the ejected common  envelope, that
once was a  planetary nebula. This ``SN  Ia Inside a PN''  is termed a
SNIP \citep{TsebrenkoSoker2015a, TsebrenkoSoker2015b}.

\subsection{The temperature of the AGB core}

For the sake  of completeness for each of the  cases described earlier
we  perform two  simulations. For  the first  of these  simulations we
adopt  a low  temperature  for the  isothermal core  of  the AGB  star
($T=10^6$~K),   while   for  the   second   we   adopt  a   hot   core
($T=10^8$~K). Admittedly, these are  first order approximations, since
realistic temperature profiles should  be adopted. However, given that
the  degeneracy of  typical  AGB  cores is  high  we  expect that  the
influence of the adopted temperature  profiles on the evolution of the
merger will be  small. The white dwarf  is always cold, and  for it we
adopt  $T=10^6$ K.   The  adopted  temperature has  an  effect on  the
initial configuration  of the core  of the  AGB star (of  mass $M_{\rm
core}=0.77\,M_{\sun}$).  The radii  are $R_{\rm core}\simeq 1.02\times
10^{-2}\,R_{\sun}$     and      $R_{\rm     core}\simeq     9.33\times
10^{-3}\,R_{\sun}$, and  the central densities are  $\rho_{\rm c}^{\rm
core}\simeq  6.83\times   10^6$~g~cm$^{-3}$  and   $\rho_{\rm  c}^{\rm
core}\simeq 7.07\times 10^6$~g~cm$^{-3}$, for  the hot and cold cores,
respectively.  Note that these differences  are small, of the order of
a  few percents,  and thus  given the  approximations adopted  in this
work, have limited  effects on the dynamical evolution  of the merger,
and  on the  question  of whether  there is  an  ignition upon  merger
\citep{Yoon2007}.

\section{Numerical setup}
\label{sec:numerical}

\begin{figure*}
   \begin{center}
   {\includegraphics[width=0.75\textwidth, angle=-90]{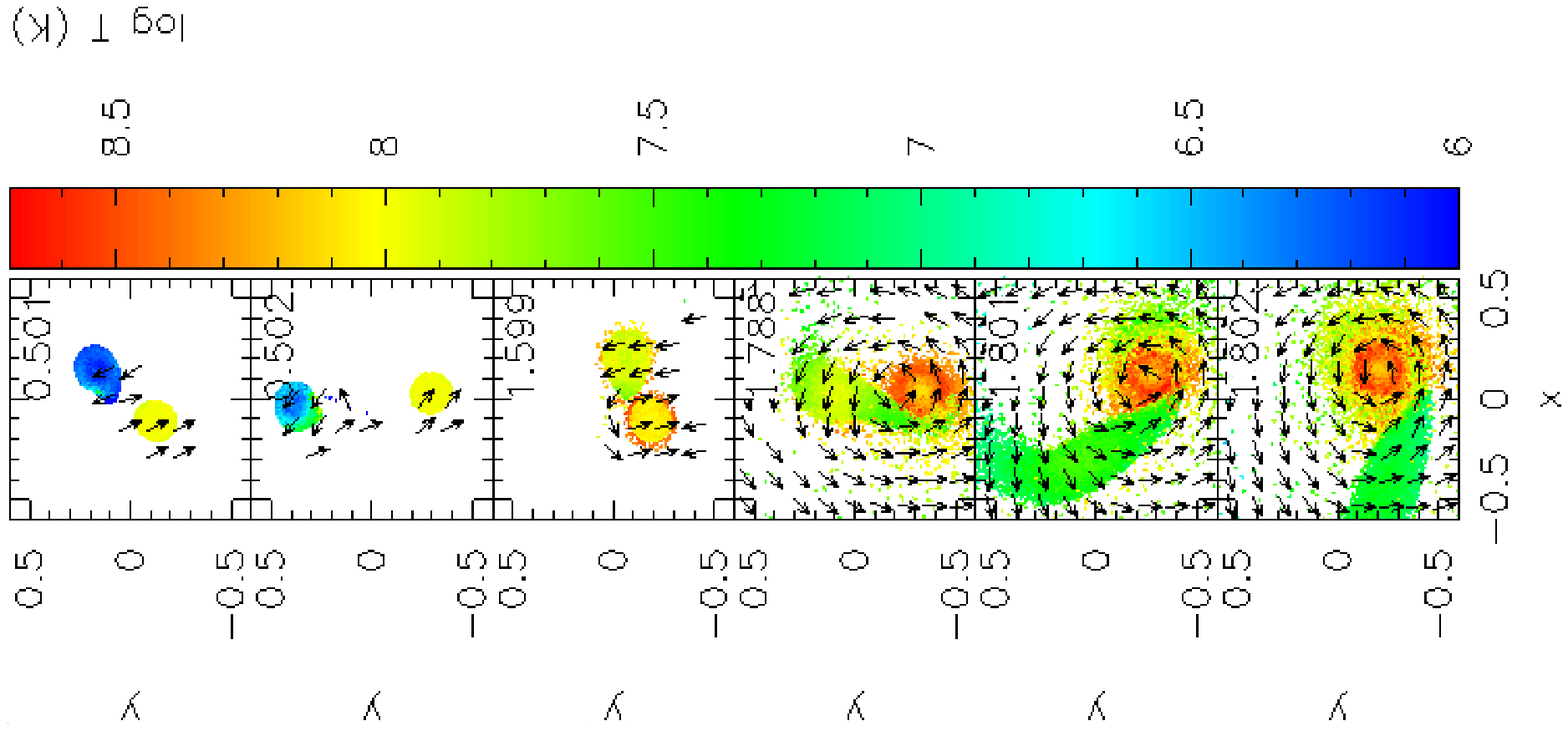}\hspace{0.2cm}
    \includegraphics[width=0.75\textwidth, angle=-90]{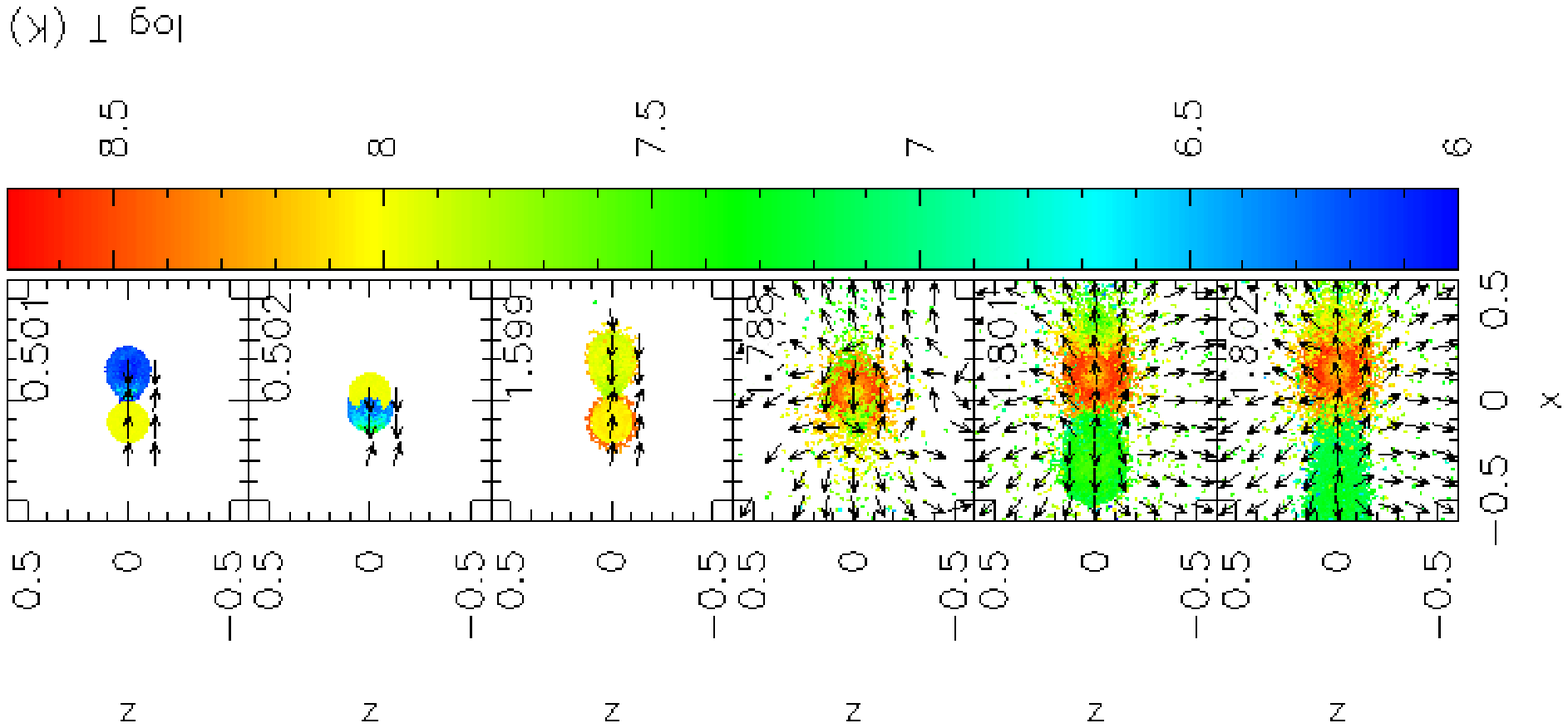}} \\
    \caption{Evolution of the binary system for selected stages of the
      merger, as described in the  main text, in the equatorial (left)
      and meridional (right) planes, for the  case in which a hot core
      of the  AGB star  is adopted and  for $q_{\rm  disk}=0.12$.  The
      color scale shows the logarithm  of the temperature, whereas the
      $x$, $y$  and $z$ axes  are in  units of $0.1\,  R_{\sun}$.  The
      arrows show  the direction  of the velocity  field, but  not its
      magnitude, in  the corresponding plane.  Each  panel is labelled
      with  the corresponding  time  in units  of  the initial  binary
      period, $T_0=10,126$~s.   The white dwarf is  the object located
      initially at the  right, and later is  destroyed.  These figures
      have   been   done   using   the   visualization   tool   SPLASH
      \citep{Price2007b}. See the online edition  of the journal for a
      color version of this figure.}
\label{fig:time12}
   \end{center}
\end{figure*}

Both the white  dwarf and the core  of the AGB star are  made of equal
mass fractions  of carbon  and oxygen, with  flat profiles.   We would
like  to mention  here that  the outer  helium layer  of $\sim  0.01\,
M_{\sun}$ surrounding the cores of the white dwarf and of the AGB star
could   be   potentially   important   for  the   evolution   of   the
merger.  However, with  the number  of  particles adopted  in our  SPH
simulations  this  helium shell  it  is  practically unresolvable  and
difficult to follow its evolution numerically.

The SPH code used in this work  is an optimized and updated version of
the code  used by \cite{Gea04} and  later by \cite{Loren-Aguilar2009},
\cite{Loren-Aguilar2010}   and  \cite{Aznar-Siguan2013}   to  simulate
double-degenerate mergers.   This code has been  extensively described
in those  works.  Consequently, here  we will only summarize  its more
important characteristics.  

\begin{figure*}
   \begin{center}
    {\includegraphics[width=0.75\textwidth, angle=-90]{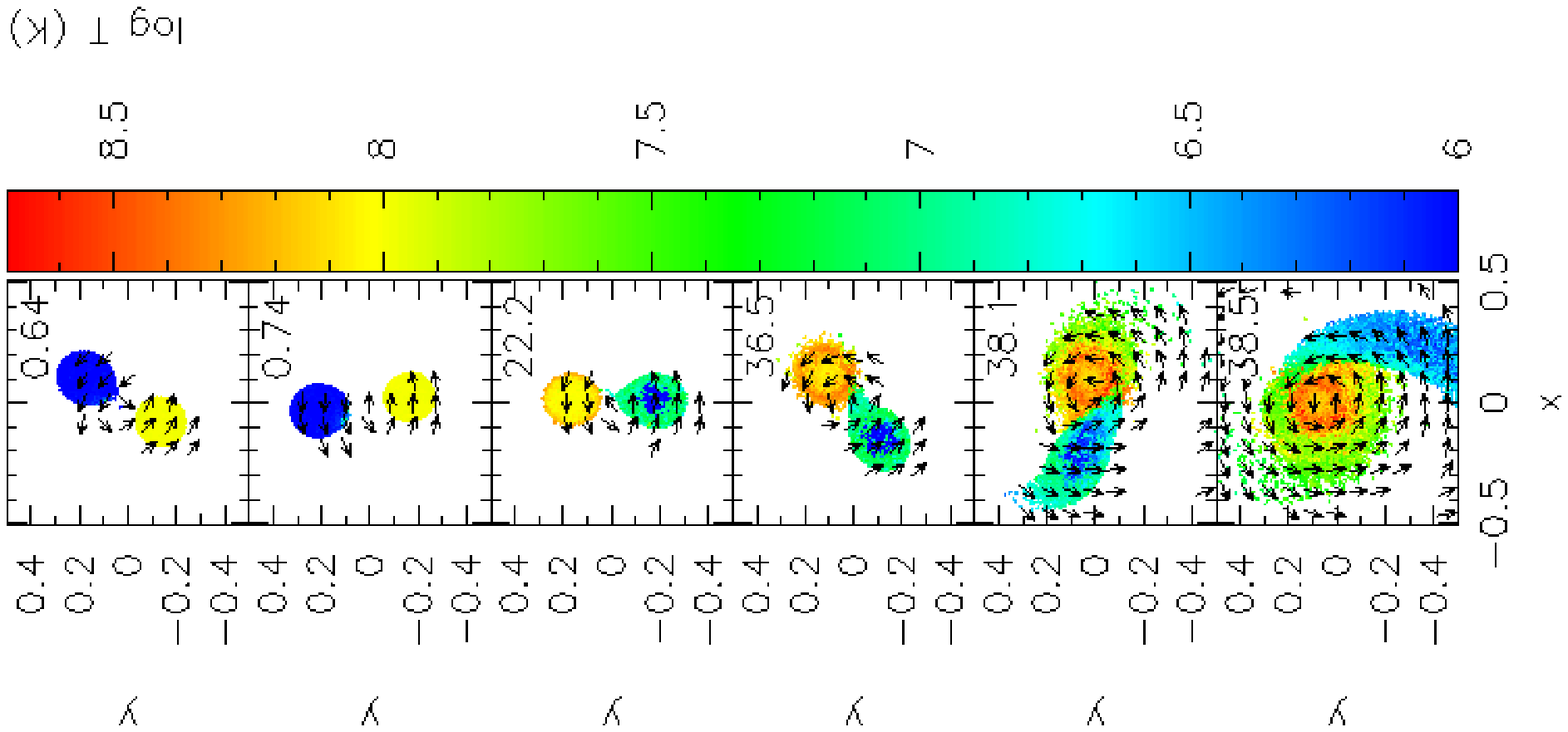}\hspace{0.2cm}
     \includegraphics[width=0.75\textwidth, angle=-90]{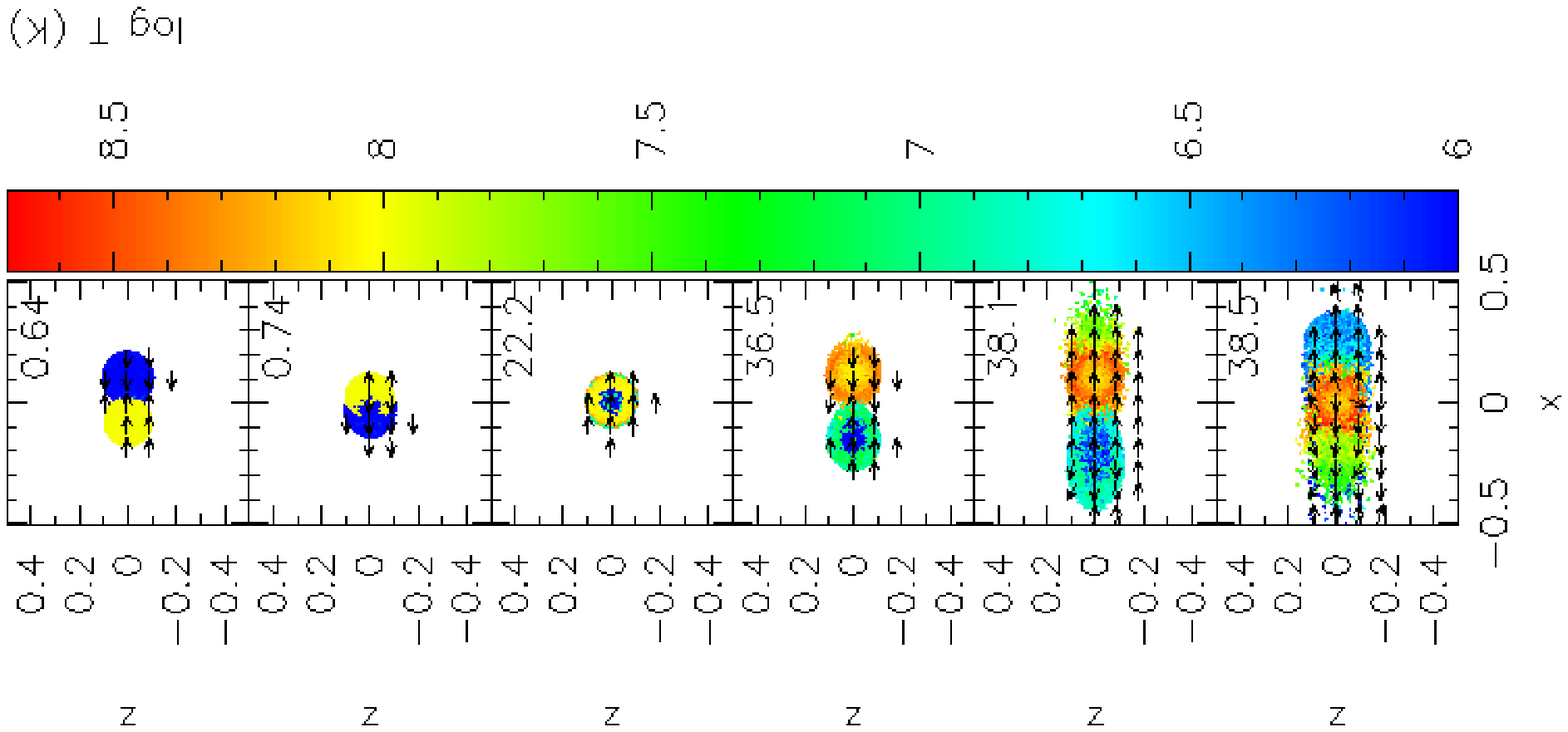}} \\
     \caption{Same   as   Fig.~\ref{fig:time12},   but   for   $q_{\rm
       disk}=0.10$.    In this  case  the initial  orbital period  is
       $T_0=65$~s.  See the online edition  of the journal for a color
       version of this figure.}
\label{fig:time10}
   \end{center}
\end{figure*}

Our SPH code  is fully parallel, and allows to  run simulations with a
large number  of particles in a  reasonable time. We use  the standard
cubic spline kernel of  \cite{ML85}.  The gravitational interaction is
also softened using this kernel \citep{HK89}.  The search of neighbors
and the evaluation of the  gravitational forces are performed using an
octree  \citep{BaHu86}. The  iteration method  of \cite{Price2007}  is
used to compute the smoothing lengths.   To deal with shocks we employ
a prescription for  the artificial viscosity based  in Riemann solvers
\citep{Monaghan1997},  and   the  variable  viscosity   parameters  of
\cite{Morris1997} are  employed. Additionally, to  suppress artificial
viscosity  forces  in  pure  shear   flows  the  viscosity  switch  of
\cite{Ba95}   is    also   used.    As   explained    in   detail   in
\cite{Aznar-Siguan2013}, we follow the  evolution of both the internal
energy and  of the temperature to  ensure a reliable evolution  of the
temperature  in a  degenerate  gas.  The  thermodynamic properties  of
matter   are  computed   using   the  Helmholtz   equation  of   state
\citep{Timmes2000}.   Finally,   the  nuclear  network   adopted  here
incorporates 14  nuclei.  The  reactions considered  in this  work are
captures of $\alpha$ particles, and the associated back reactions, the
fusion of two carbon nuclei and the reaction between carbon and oxygen
nuclei.   All the  thermonuclear reaction  rates were  taken from  the
REACLIB  data  base  \citep{Cyburt2010}.  Neutrino  losses  were  also
included, being the corresponding rates those of \cite{itoh}.

\begin{figure*}
   \begin{center}
   {\includegraphics[width=0.47\textwidth, trim=0.1cm 0.1cm 0.1cm 0.1cm,clip=true]{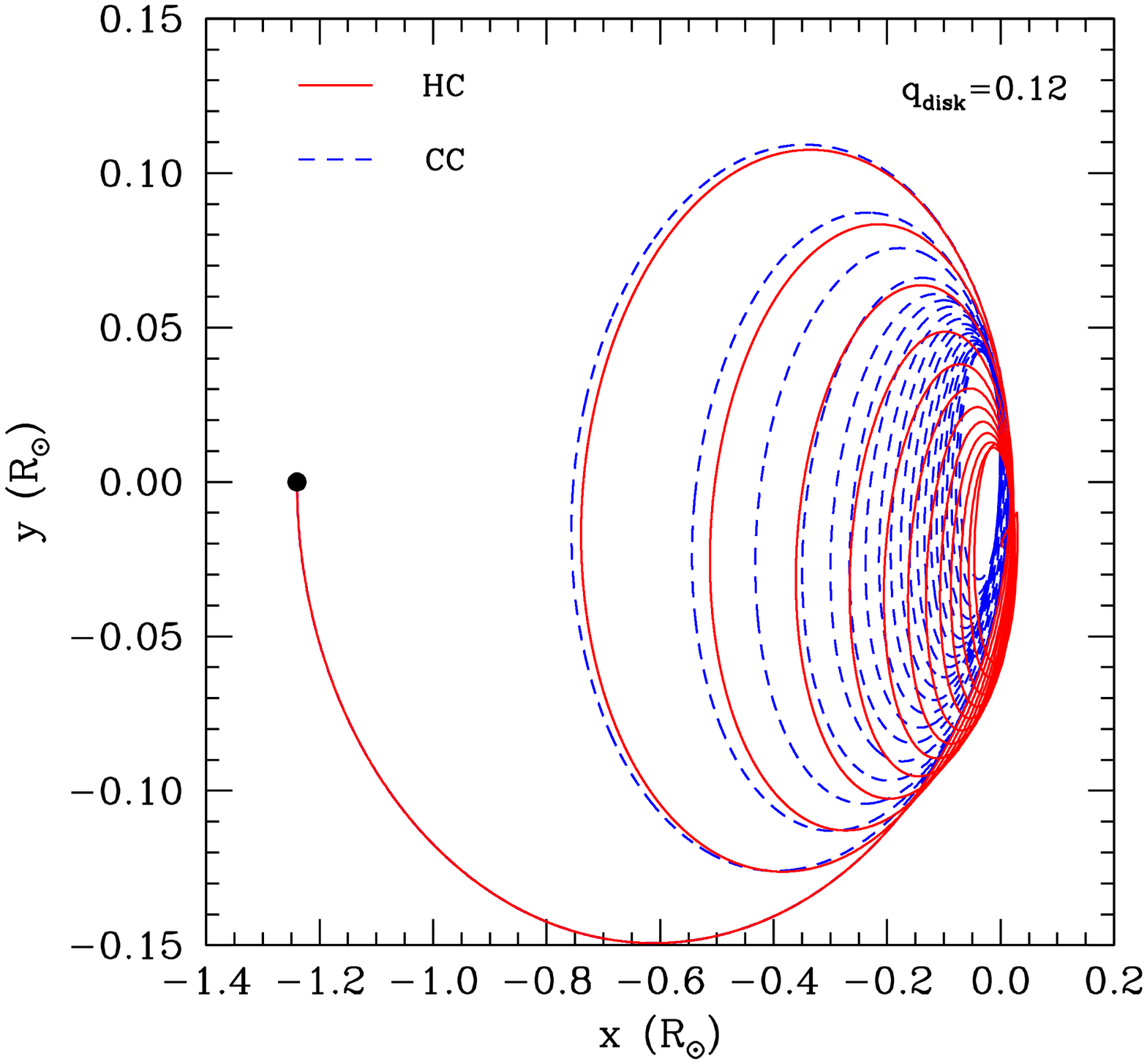}\hspace{0.3cm}
    \includegraphics[width=0.47\textwidth, trim=0.1cm 0.1cm 0.1cm 0.1cm,clip=true]{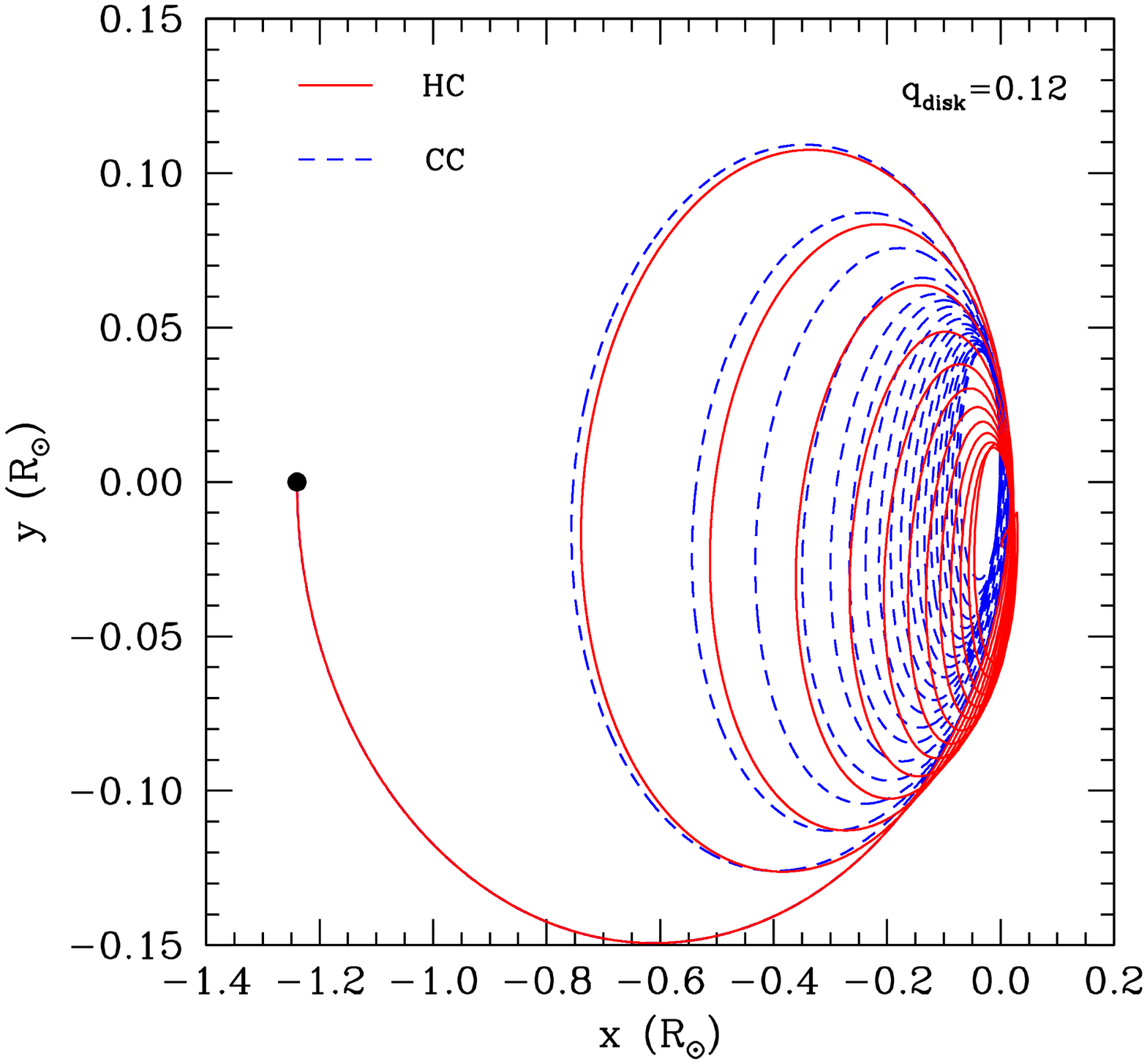}}\\
   {\includegraphics[width=0.47\textwidth, trim=0.1cm 0.1cm 0.1cm 0.1cm,clip=true]{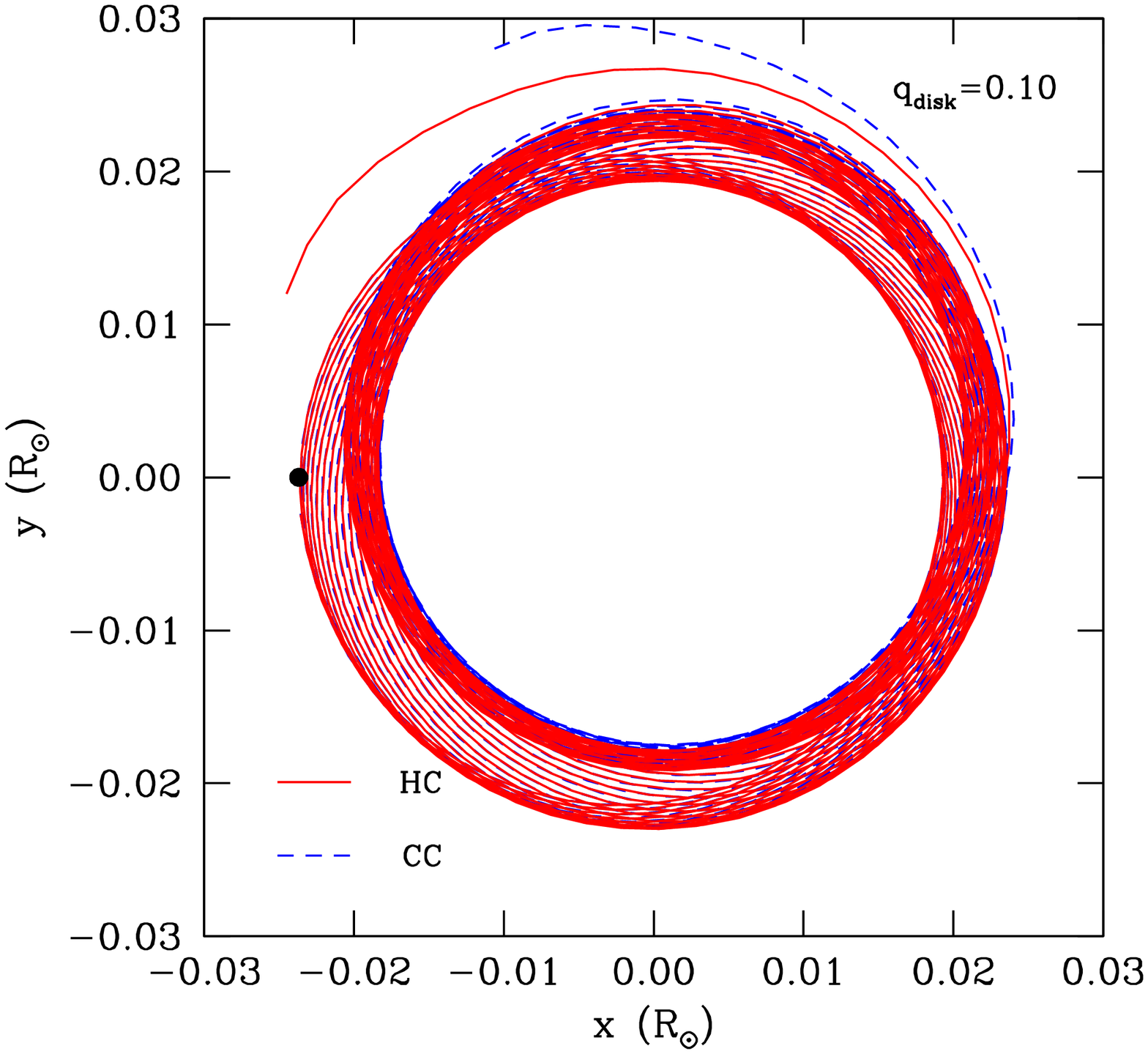}\hspace{0.3cm}
    \includegraphics[width=0.47\textwidth, trim=0.1cm 0.1cm 0.1cm 0.1cm,clip=true]{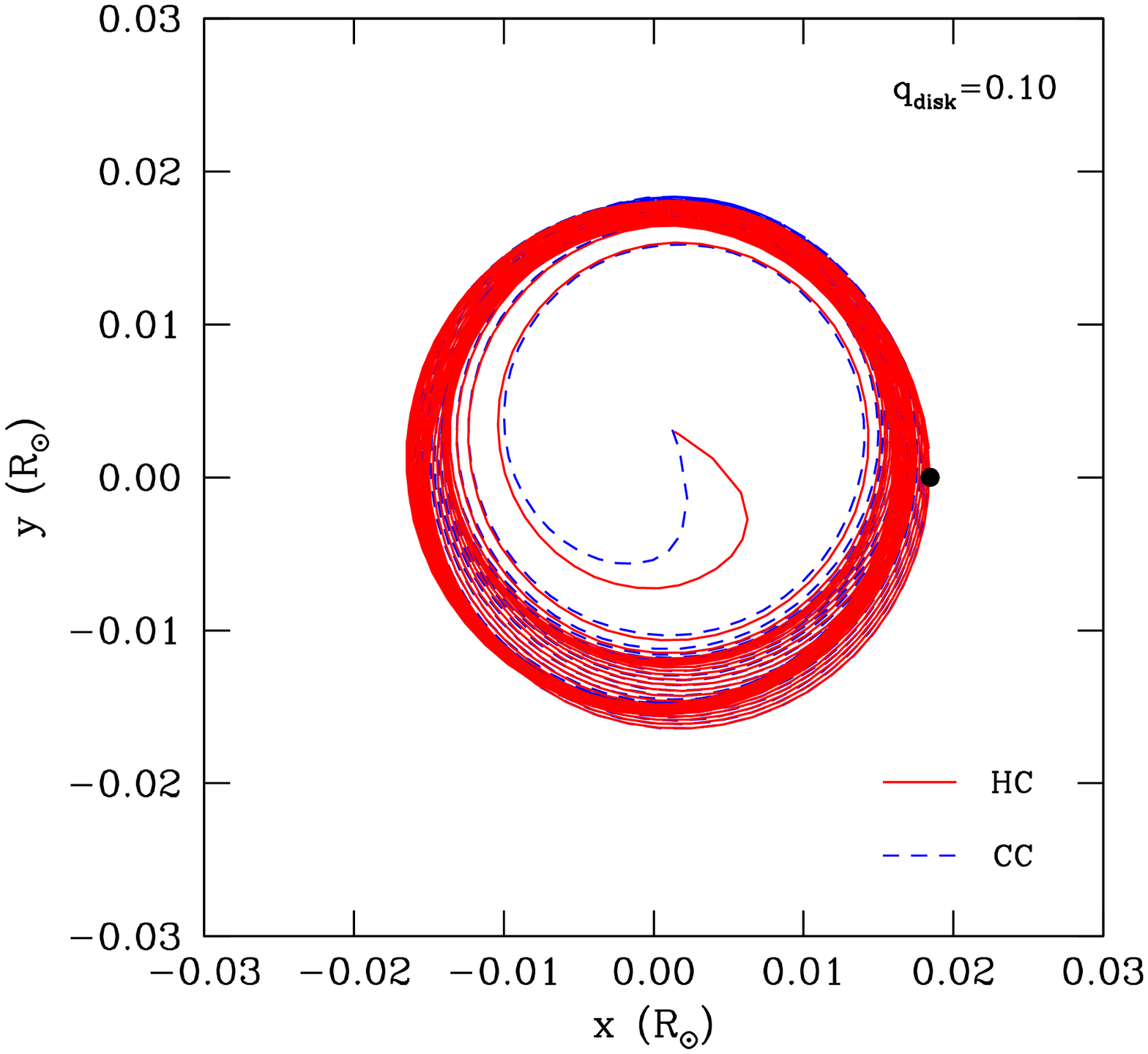}}\\
   \caption{Trajectories of  the centers of  mass of the  white dwarfs
     (left panel)  and cores of  the AGB  stars (right panel)  for the
     case in  which $q_{\rm  disk}=0.12$ is  adopted (top  panels) and
     that in  which $q_{\rm  disk}=0.10$ is employed  (bottom panels).
     The blue dashed lines correspond to  the case in which a cold AGB
     core is adopted, and  the red solid lines to that  in which a hot
     core is  assumed. The  solid black  circles indicate  the initial
     positions of both stars for each of the panels.}
\label{fig:orb}  
   \end{center}
\end{figure*}

As  mentioned before,  the  first mass  transfer  episode begins  when
$r_{\rm  min}<r_{\rm O}<a$.   To obtain  reliable configurations  when
both stars are at closest approach  we proceeded as follows.  We first
relaxed independent configurations for each of the stars of the binary
system and we placed them at the apastron in a counterclockwise orbit.
We  then  evolved  the  system  in  the  corotating  frame  to  obtain
equilibrium  configurations at  this distance.   Once this  relaxation
process  was finished,  we  started the  simulations  in the  inertial
reference  frame.   In  a  first set  of  preliminary  simulations  we
explored at which  distance the mass transfer episode  ensues. This is
done employing a reduced number of particles ($\sim 2 \times 10^4$ for
each star).   Once we know  which are  the orbital separation  and the
eccentricity of the orbit for which we obtain a Roche-lobe overflow we
computed a second  set of simulations with  enhanced resolution. Since
for highly eccentric orbits both stars are initially separated by very
large  distance,  to  save  computing  time  in  this  second  set  of
simulations we  first followed the  evolution of the  system employing
the reduced resolution  and, once the stars had  completed one quarter
of the orbit,  these low-resolution simulations were  stopped and both
stars were remapped using a large number of SPH particles (a factor of
10  larger).   We then  resumed  the  simulations with  this  enhanced
resolution.    This  was   done   because  for   systems  with   large
eccentricities the computing load  required to follow an uninteresting
phase of the orbital evolution of the pair is exceedingly large.

\begin{figure*}
   \begin{center}
   {\includegraphics[width=0.47\textwidth, trim=0.1cm 0.1cm 0.1cm 0.1cm,clip=true]{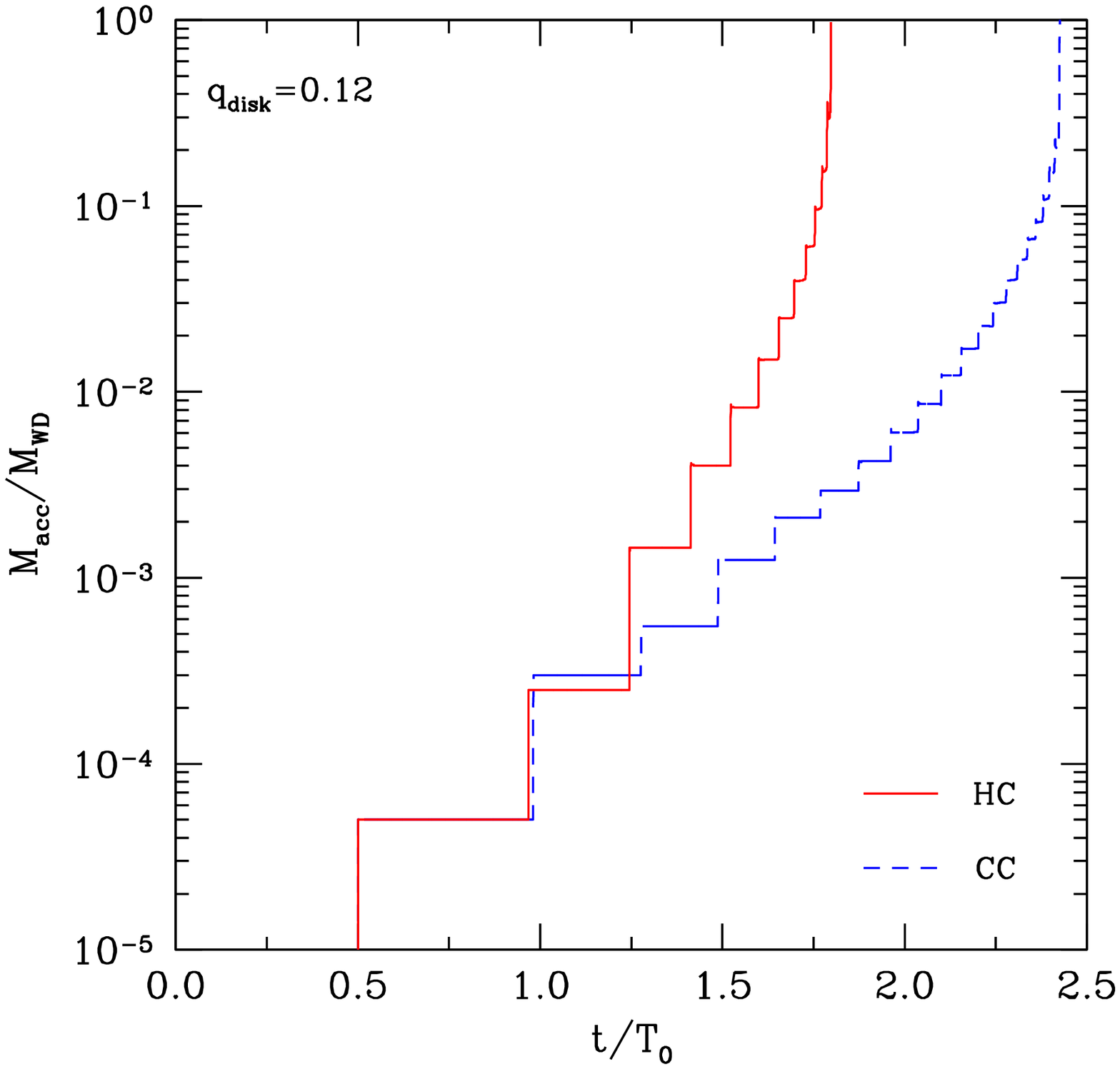}\hspace{0.3cm}
    \includegraphics[width=0.47\textwidth, trim=0.1cm 0.1cm 0.1cm 0.1cm,clip=true]{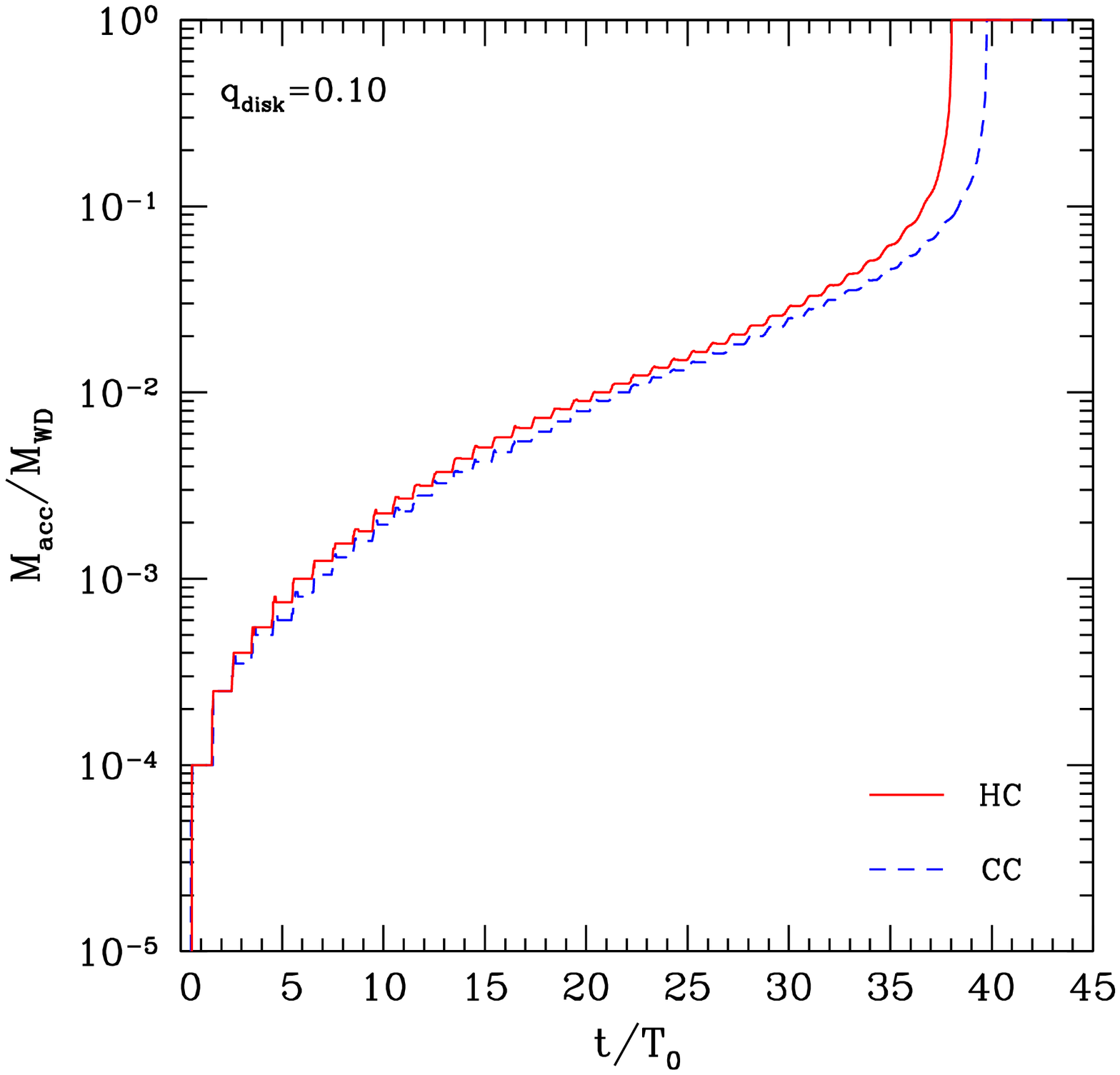}}\\
   \caption{Mass  accreted by  the AGB  core  as a  function of  time,
     expressed in terms of the initial orbital period.  The left panel
     corresponds to $q_{\rm disk}=0.12$,  while the right displays the
     case in  which $q_{\rm  disk}=0.10$.  Each  step in  these panels
     corresponds  to a  periastron  passage. The  total  time span  is
     $\simeq 25,315$~s  for the left  panel, and $\simeq  2,925$~s for
     the right one. }
\label{fig:macc}  
   \end{center}
\end{figure*}
 
\section{Results}
\label{sec:results}

As  discussed  earlier,  in  this paper  we  simulate  four  different
configurations for  the binary system.  In all  cases the mass  of the
core of  the AGB star and  the white dwarf are,  respectively, $0.77\,
M_{\sun}$ and $0.60\, M_{\sun}$.   The first configuration corresponds
to a  binary system that  has a  disk with a  mass ratio equal  to the
critical  one of  $q_{\rm disk}=0.12$  (see equation  \ref{eq:qdisk}),
while for the  second one we have  chosen a disk with a  mass ratio of
$q_{\rm disk}=0.1$, which is below the critical value. For each one of
these cases  we have studied  the coalescence when the  temperature of
the core of the AGB star is high ($T=10^8$~K) and low ($T=10^6$~K).

For the highly eccentric orbit that results when $q_{\rm disk} = 0.12$
is adopted, the  first mass transfer from the white  dwarf to the core
of the  AGB star occurs when  the distance between both  components of
the  binary system  at closest  approach is  $r_{\rm min}=0.90\,a_{\rm
ECO}$,  where $a_{\rm  ECO}$ has  been defined  previously in  section
\ref{CE} as the  radius of a circular orbit for  which the white dwarf
overflows  its Roche  lobe.  In  computing this  orbital distance  the
classical analytical expresion of \cite{Eggleton83} for the Roche lobe
radius $R_{\rm L}$  was used. Note that this expression  does not take
into  account  tidal deformations,  and  is  only valid  for  circular
orbits,  while the  orbits  of  the binary  systems  studied here  are
elliptical  and our  stars are  tidally distorted.   At this  time the
semimajor axis is $a=1.12\,R_{\sun}$ and the eccentricity is $e=0.97$.
For the less  massive disk with $q_{\rm disk}=0.10$  the merger begins
when $a=0.038\,R_{\sun}$  and the eccentricity  of the orbit  is small
$e=0.095$.    Specifically,    the   merger   occurs    when   $r_{\rm
min}=0.98\,a_{\rm  ECO}$.   Fig~\ref{fig:ic}   illustrates  all  this.
Particularly relevant for the discussion of our results are the values
of the  respective initial periods.  For  the case of the  system with
$q_{\rm  disk}=0.12$   the  initial   orbital  period   is  $T_0\simeq
10,126$~s, while for the case in which $q_{\rm disk}=0.10$ is used the
initial period is significantly shorter, $T_0\simeq 65$~s.

\subsection{Evolution of the merger}

Fig.~\ref{fig:time12} shows  the time  evolution of the  binary system
during the  merger for the  case in which  the temperature of  the AGB
core is $10^8$~K and for  $q_{\rm disk}=0.12$. Shown are the positions
of  the SPH  particles and  their color-coded  temperatures. The  left
column shows the  evolution in the orbital plane and  the right column
in  the meridional  plane, respectively.  In Fig.~\ref{fig:time10}  we
present the results in the same  setting for the case in which $q_{\rm
disk}=0.10$  is  adopted.   The  white dwarf  is  the  object  located
initially at the right, and later  is destroyed, while the core of the
AGB star is initially located at the left.

\begin{figure*}
   \begin{center}
   {\includegraphics[width=0.47\textwidth, trim=0.1cm 0.1cm 0.1cm 0.1cm,clip=true]{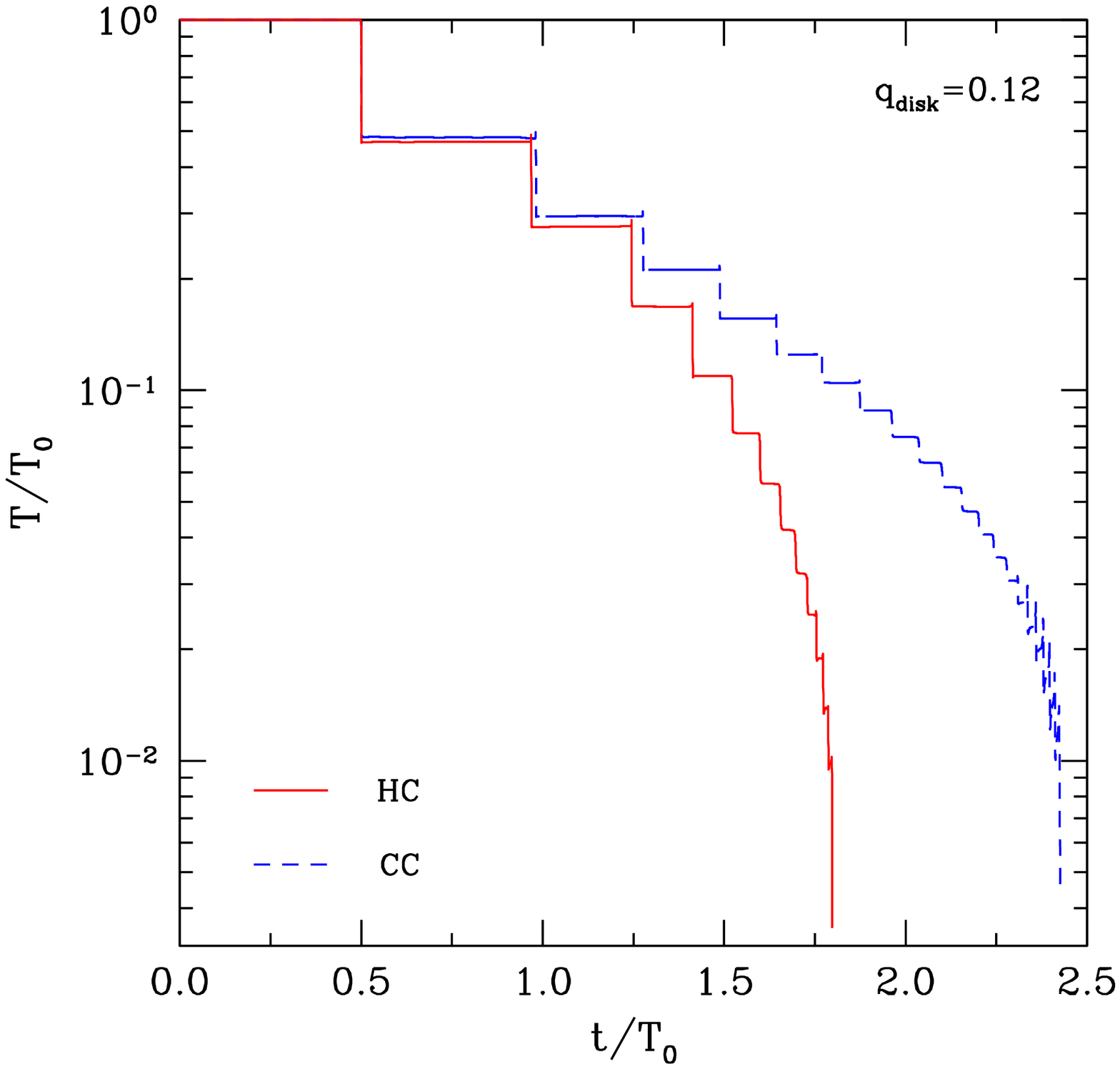}\hspace{0.3cm}
    \includegraphics[width=0.47\textwidth, trim=0.1cm 0.1cm 0.1cm 0.1cm,clip=true]{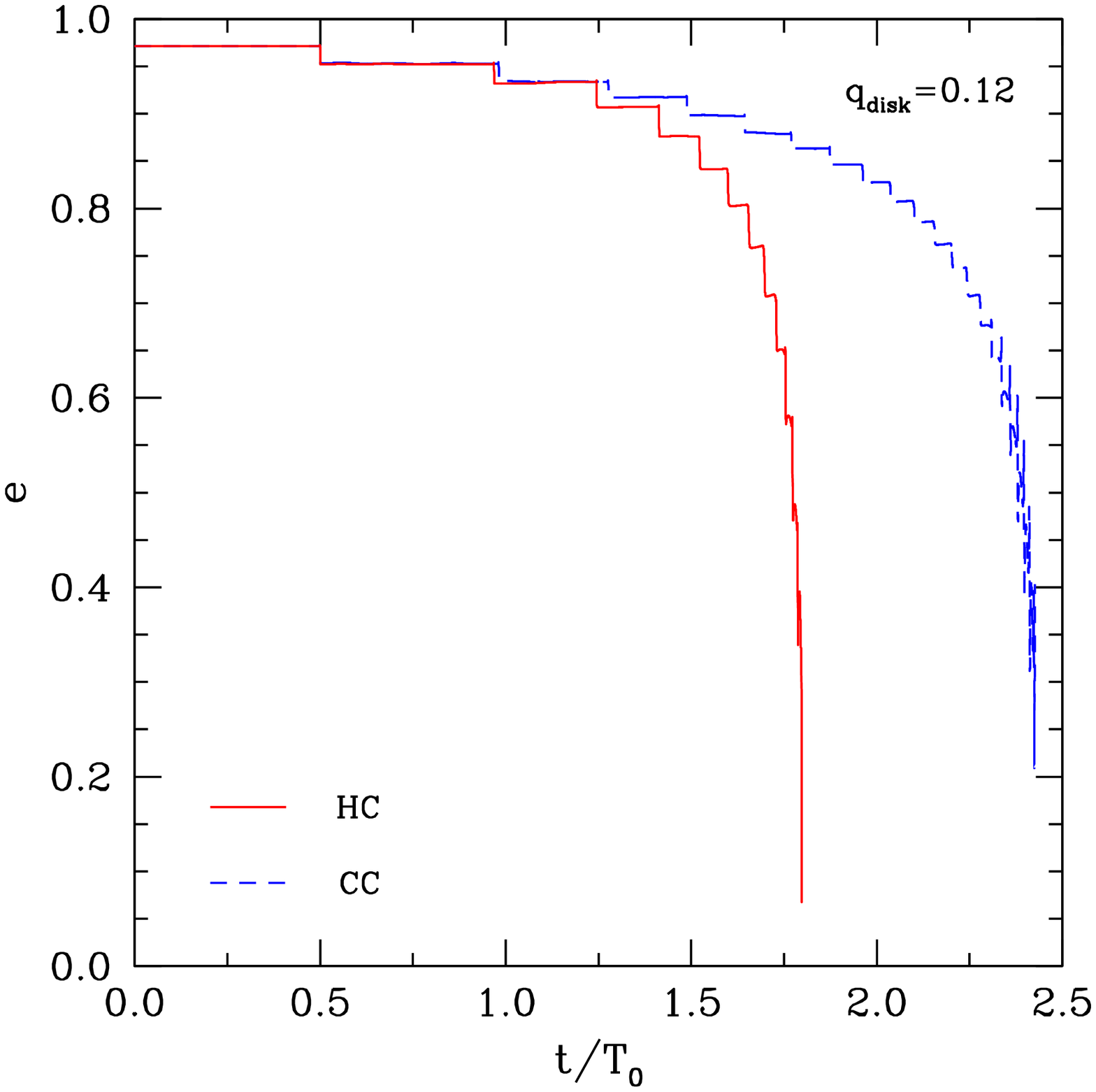}}\\
   {\includegraphics[width=0.47\textwidth, trim=0.1cm 0.1cm 0.1cm 0.1cm,clip=true]{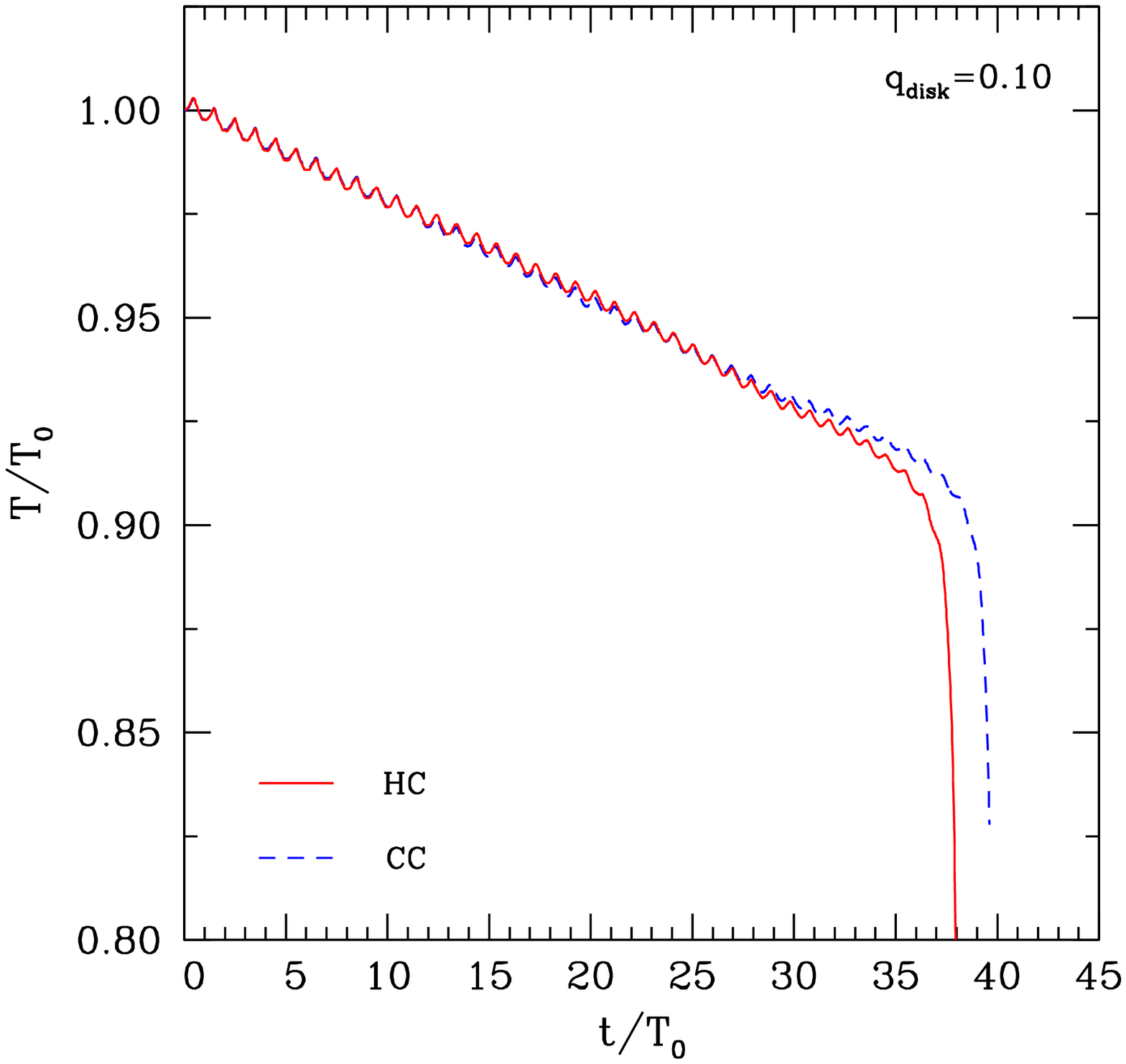}\hspace{0.3cm}
    \includegraphics[width=0.47\textwidth, trim=0.1cm 0.1cm 0.1cm 0.1cm,clip=true]{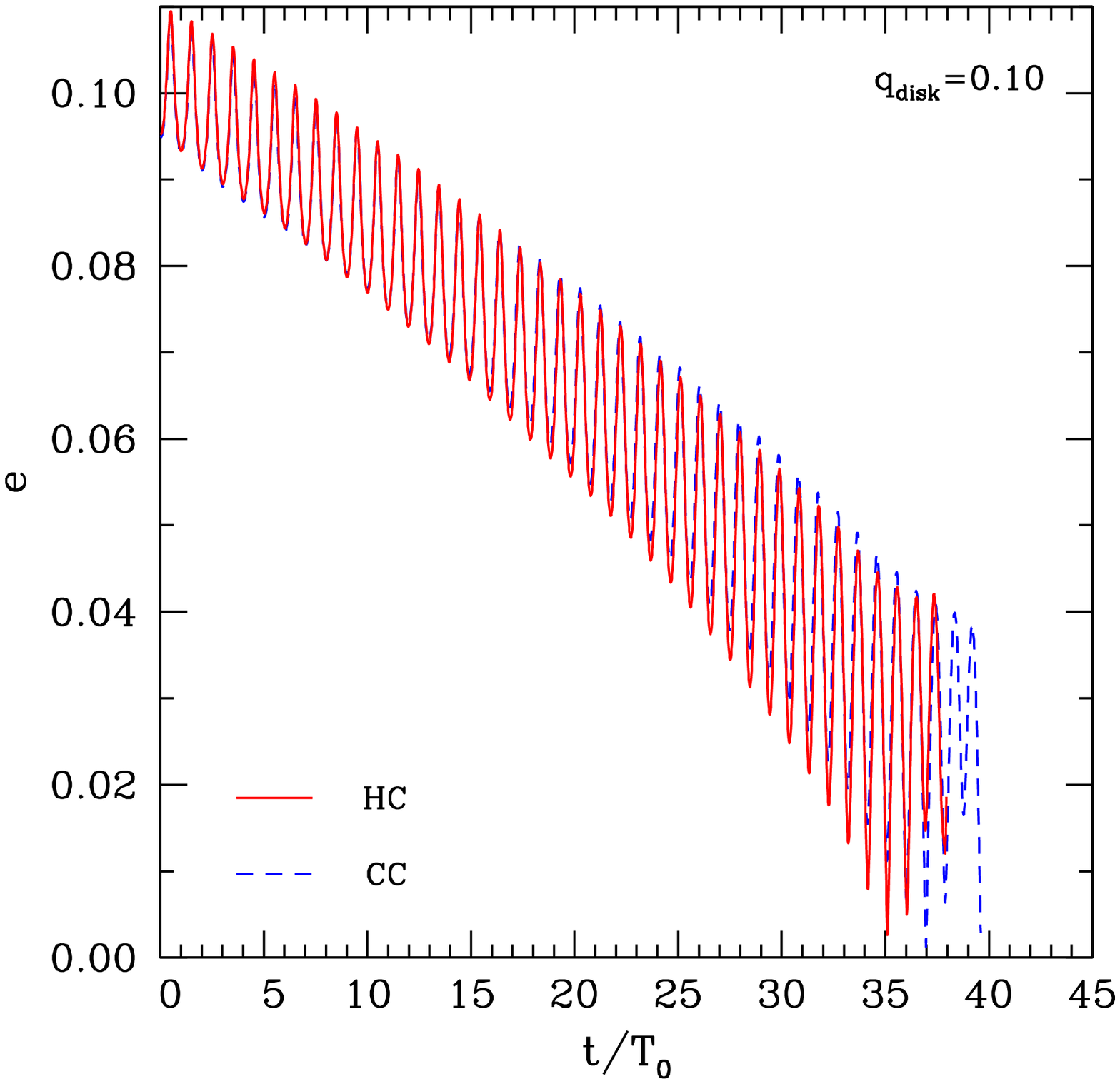}}\\
   \caption{The  evolution of  the periods  (left panels)  and of  the
     eccentricities (right panels) during the merger process.  The top
     panels  represent  the  case  in  which  $q_{\rm  disk}=0.12$  is
     considered,  and   the  bottom   ones  those  in   which  $q_{\rm
     disk}=0.10$ is adopted.}
\label{fig:params} 
   \end{center}
\end{figure*}

The top  panels of Figs.~\ref{fig:time12} and  \ref{fig:time10} are at
the initial  stages of  the merger, at  times $t/T_0\simeq  0.501$ for
$q_{\rm disk}=0.12$, and $t/T_0\simeq  0.640$ for $q_{\rm disk}=0.10$,
respectively. We  remind that  the orbital periods  are $T_0=10,126$~s
for  $q_{\rm  disk}=0.12$,  and $T_0=65$~s  for  $q_{\rm  disk}=0.10$,
respectively.   As  can  be  seen the  white  dwarf  is  significantly
deformed, due to tidal interactions,  the deformation being larger for
the  case  of an  eccentric  orbit  (Fig.~\ref{fig:time12}).  For  the
second  row   of  panels  we   have  chosen  slightly   larger  times,
$t/T_0\simeq  0.502$ and  $t/T_0\simeq 0.74$,  respectively, and  show
that  tidal  deformations close  to  periastron  are much  larger  for
eccentric  mergers. This  leads to  a faster  heating of  the external
layers of the secondary star. The third rows show the systems when the
system  has evolved  through several  passages trough  the periastron.
For  $q_{\rm disk}=0.12$  (Fig.~\ref{fig:time12}) we  have chosen  the
sixth   mass   transfer   episode    and   for   $q_{\rm   disk}=0.10$
(Fig.~\ref{fig:time10})  the  twenty-third  one. These  correspond  to
times $t/T_0\simeq 1.599$ and $t/T_0\simeq 22.2$, respectively. As can
be seen, at these evolutionary stages in both cases the secondary (the
white dwarf)  has increased  its temperature, and  more mass  has been
accumulated on the primary core.

The  fourth rows  of Figs.~\ref{fig:time12}  and \ref{fig:time10},  at
times $t/T_0\simeq 1.788$ and $t/T_0\simeq 36.5$ respectively, display
the situation during  the last orbit, just before  the secondary white
dwarfs are completely disrupted by the cores of the AGB stars. At this
point, for  $q_{\rm disk}=0.12$ the  white dwarf still  keeps $\approx
80\%$  of its  initial mass,  while  for $q=0.10$  this percentage  is
somewhat larger, $\approx 90\%$.  The  fifth rows correspond to a time
close to that at which the  peak temperature is reached, which happens
during the final infall of the secondary star onto the core of the AGB
star.    Specifically,  we   have  chosen   $t/T_0\simeq  1.801$   and
$t/T_0\simeq 38.1$,  respectively.  As  it will  be discussed  in more
detail later,  for $q_{\rm  disk}=0.12$ these temperatures  are higher
than for $q_{\rm disk}=0.10$.  The  reason for this is twofold. First,
the material of  the white dwarf is hotter for  $q_{\rm disk}=0.12$, a
consequence of the  previous evolution.  The second reason  is that in
this  case the  matter of  the disrupted  white dwarf  falls onto  the
primary  star with  a much  larger  radial velocity  than for  $q_{\rm
disk}=0.10$. Consequently, matter is compressed more violently in this
case. Finally, the last rows show  the systems at late times.  We have
chosen times $t/T_0\simeq 1.802$ and $t/T_0\simeq 38.5$, respectively.
As can  be seen,  the material of  the completely  disrupted secondary
white dwarf is spiralling around the primary star.

\begin{figure*}
   \begin{center}
   {\includegraphics[width=0.47\textwidth, trim=0.1cm 0.1cm 0.1cm 0.1cm,clip=true]{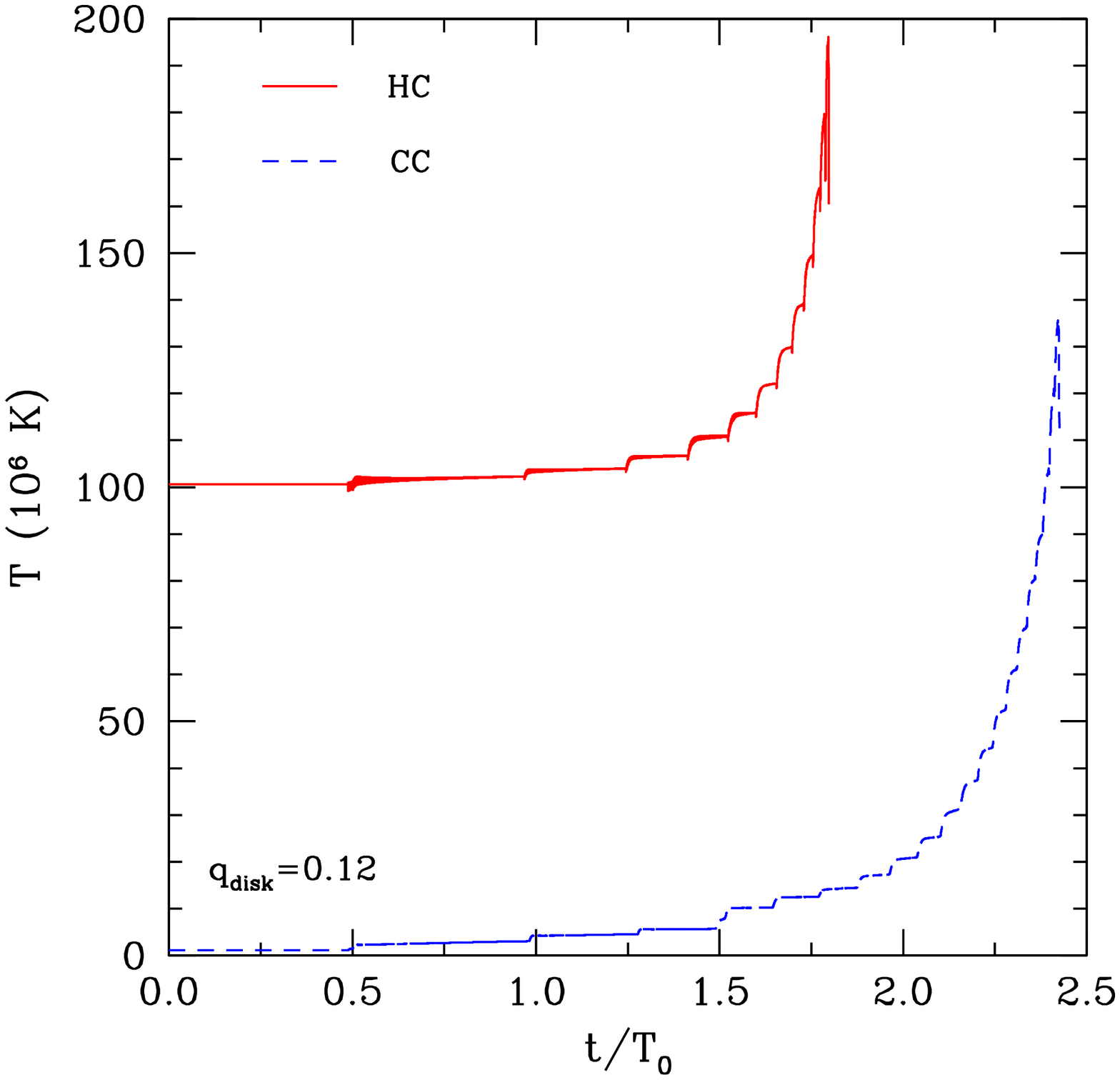}\hspace{0.3cm}
    \includegraphics[width=0.47\textwidth, trim=0.1cm 0.1cm 0.1cm 0.1cm,clip=true]{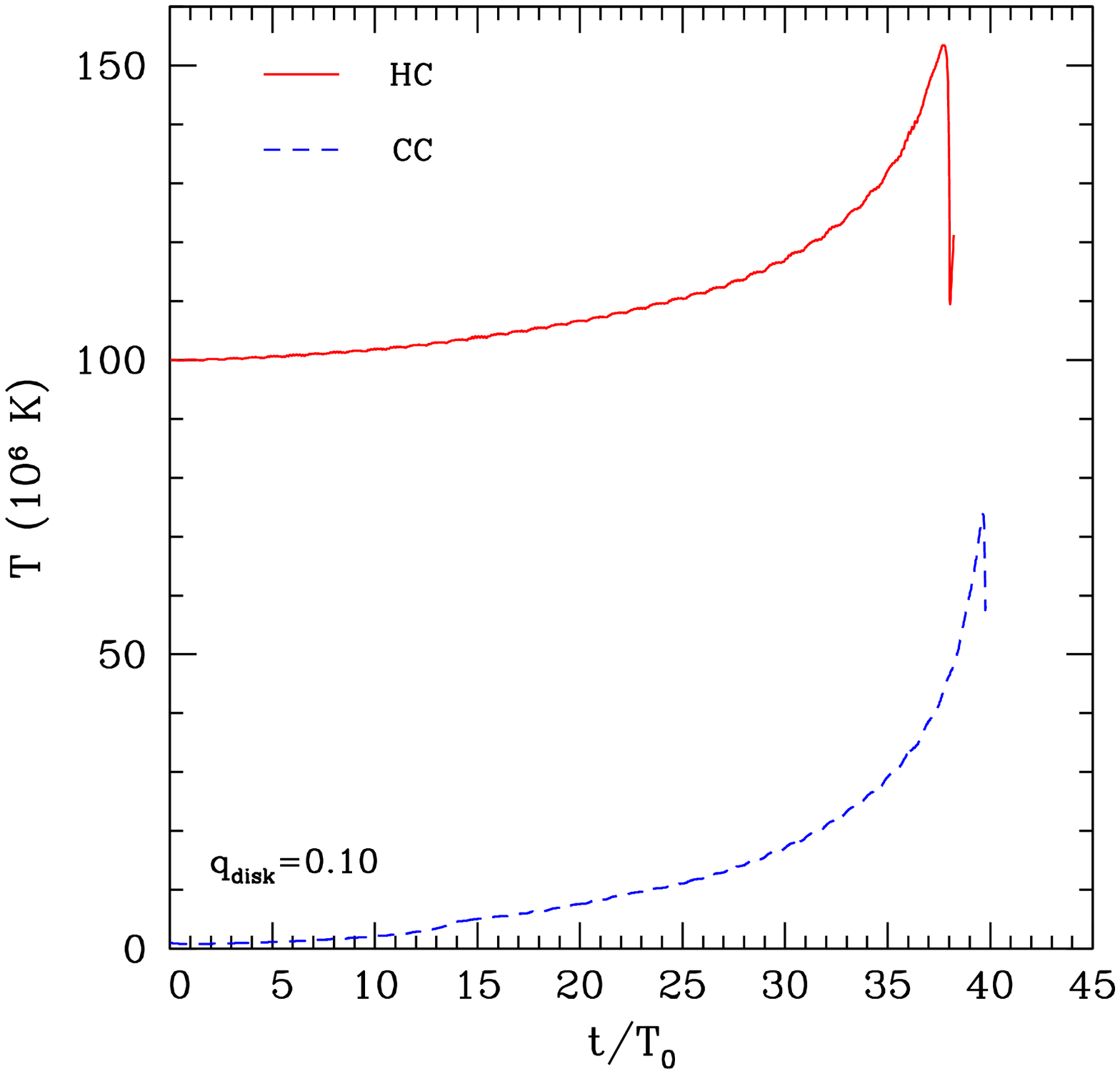}}\\
   \caption{Average temperature of the core of the AGB star during the
     merger, for the  case in which $q_{\rm  disk}=0.12$ is considered
     (left panel),  and for that  in which $q_{\rm  disk}=0.10$ (right
     panel).}
\label{fig:tem1}
   \end{center}
\end{figure*}

A general feature  of our simulations is that the  secondary star, the
white   dwarf,  is   totally   disrupted   after  several   periastron
passages. In each of these  periastron passages mass is transferred in
a relatively  gentle way.  Accordingly,  in all cases the  density and
temperature conditions for a detonation to develop are not met. As can
be seen in Figs.~\ref{fig:time12} and \ref{fig:time10}, in none of the
cases  the  peak temperature  reaches  $10^9$~K,  and hence,  although
nuclear reactions  play a role,  a powerful explosion does  not occur.
Nevertheless, the helium shells of the white dwarf and the core of the
AGB star can  play a role in inducing a  detonation.  However, none of
the   studies    of   merging   white   dwarfs    performed   so   far
\citep{Guillochon_2010,   Townsley_2012,    Moore_2013,   Pakmor_2013,
Shen_2014} has been  able to self-consistently evolve  a binary system
with  stable helium  layers, and  consequently this  is a  topic that,
although interesting,  deserves further  scrutiny.  This  is, however,
beyond the scope  of this paper.  Another point of  concern is how the
resolution (that  is, the  number of SPH  particles) affects  the peak
temperature.  This has been investigated in several works --- see, for
instance, the discussions about this issue in \cite{Aznar-Siguan2013},
and  references  therein.   It  is   typically  found  that  the  peak
temperatures may differ by a factor of 2 when low- and high-resolution
simulations  are  compared, but  these  temperatures  do not  show  an
excessive dependence on the adopted  resolution when the number of SPH
particles is  larger than  $\sim 10^5$  --- typically  the differences
amount to $\sim 15\%$.

We emphasize that when $q_{\rm disk}=0.12$ is adopted --- that is, for
those mergers  for which the  eccentricity is high ---  the successive
mass  transfer  episodes   only  happen  when  both   stars  are  very
close. Actually,  it turns  out that  when both  stars are  at closest
approach the secondary white  dwarf is substantially tidally deformed,
and the  mass transfer episode  occurs just after the  passage through
the periastron.   However, soon after this,  tidal deformations become
smaller, so  the mass transfer  episode stops, and both  components of
the binary  system recover  their initial  spherical symmetry  for the
rest  of the  orbit.  On  the  contrary, when  $q_{\rm disk}=0.10$  is
employed  ---  corresponding to  mergers  driven  by the  emission  of
gravitational waves ---  the orbits are almost  circular, and although
the mass  transfer episodes  take place also  at periastron,  when the
secondary is  significantly distorted  by tidal forces,  the secondary
star  remains tidally  deformed for  the  entire orbit,  and thus  the
merger process proceeds in a smoother way.

In all  four cases  the successive mass  transfer episodes  modify the
mass ratio of  both components of the binary system,  and this in turn
changes  the   respective  orbits.  Fig.~\ref{fig:orb}   displays  the
trajectories of  the center of mass  of both components of  the binary
system.  The left panels of this figure depict the trajectories of the
centers  of  mass of  the  white  dwarf  components.  The  blue  lines
correspond to  the cases  in which the  core of the  AGB star  is cold
(labelled as CC), while the red ones  show the cases in which we adopt
a high temperature for  the core of the AGB star  (labelled as HC). In
the right  panels the trajectories  of the center  of mass of  the AGB
cores are displayed,  using the same color coding.  The  top panels of
this figure display the case in  which $q_{\rm disk}=0.12$.  As can be
seen,  in this  case the  orbits are  eccentric, while  in the  bottom
panels, corresponding to $q_{\rm disk}=0.10$, the orbits are initially
nearly circular.   As mentioned earlier,  for both cases it  turns out
that during  the first  periastron passage  the white  dwarf transfers
mass to  the heavier AGB  core and  its radius increases.  This radius
increase starts  a series  of successive  mass transfer  episodes that
ultimately lead to a merger.

We stress that  the number of mass transfer episodes  depends not only
on the physical  characteristics of the binary system but  also on the
number of SPH  particles employed in the simulations, that  is, on the
spatial resolution.   Although our  resolution is enough  for studying
the  overall properties  of the  dynamical interaction  and the  gross
properties  of the  merged  remnants, it  is clear  as  well that  our
simulations provide only a lower limit  to the number of mass transfer
episodes.  For this reason we do not quote the number of mass transfer
episodes. Note as well that for  the case in which $q_{\rm disk}=0.10$
it  is quite  apparent  that the  last orbits  are  not circular,  but
instead during the final stages of  the merger the orbit becomes quite
eccentric.  This  final infall phase also  takes place in the  case of
eccentric mergers.  However, due  to the difference  of scales  of the
axis,   it   cannot   be   well    seen   in   the   top   panels   of
Fig.~\ref{fig:orb}.  Finally,  it  is   also  worth  noting  that  the
trajectories of  the centers of  mass for the  case in which  cold AGB
cores are considered are only  slightly different of those obtained in
the case  of hot AGB cores.  Hence, the effects of  the temperature of
the AGB core on the overall dynamical evolution are very minor.

Figure~\ref{fig:macc} shows the  mass accreted by the core  of the AGB
star  ($M_{\rm  acc}$), in  units  of  the  initial white  dwarf  mass
($M_{\rm WD}$), as  a function of time in units  of the initial period
of the binary system, $T_0$.  As in Fig.~\ref{fig:orb}, the blue lines
correspond to the case in which a cold merger is considered, while the
red ones show the evolution for the hot merger.  Again, the left panel
displays  the evolution  for  $q_{\rm disk}=0.12$,  whereas the  right
panel shows the same for $q_{\rm disk}=0.10$. The mass accreted by the
core  of the  AGB star  corresponds  to the  total mass  of those  SPH
particles originally  belonging to  the disrupted secondary  for which
the gravitational  attraction of the  core of  the AGB star  is larger
than that of the white dwarf.

Inspecting Fig.~\ref{fig:macc}  we find that the  mass accreted during
the first mass transfer episode does  not depend on the temperature of
the AGB core, but depends  critically on the adopted eccentricity.  It
is also interesting to point out that  for the case in which a hot AGB
core is adopted, mass is transferred faster than during the first mass
transfer episode  in successive  passages through the  periastron, for
both  $q_{\rm disk}=0.12$  and  $0.10$. This  is  more noticeable  for
$q_{\rm disk}=0.12$  --- i.e., a  merger in which the  eccentricity is
high. Also, it is  important to realize that for the  cases in which a
hot AGB core is adopted the amount  of mass transferred in each one of
the episodes is always  larger than for the cases in  which a cold AGB
core is considered.  All this stems from  the fact that for  a hot AGB
core the  degeneracy of the  material of  the outer layers  is smaller
than  that of  a cold  core, and  thus the  core of  the AGB  star can
accommodate more accreted mass.

In all  the four cases  studied here  the mass transferred  during the
second mass  transfer episode is  larger than that  transferred during
the first periastron passage. This is a consequence of the substantial
change  of the  orbital  period after  the  first periastron  passage,
especially for  $q_{\rm disk}=0.12$.  Specifically, for  this case the
period decreases  by $\approx 50\%$ as  seen in Fig.~\ref{fig:params}.
This is  less evident  for the  case in  which $q_{\rm  disk}=0.10$ is
adopted because in this case the period decreases by a modest $\approx
0.2\%$.  All this  translates into a smoother evolution,  also for the
successive  mass  transfer  episodes,   for  the  $q_{\rm  disk}=0.10$
cases. Thus, it turns out that  the duration of the mergers is shorter
for the cases in  which a hot AGB is adopted,  this feature being more
pronounced   for   the   case    in   which   eccentric   orbits   are
considered. Finally, note as well that in each accretion episode there
is a small decline in the accreted mass.  These declines correspond to
matter, that although being initially accreted  by the core of the AGB
star, is bounced back to the debris region shortly after.

To  better   illustrate  the   dynamical  evolution  of   the  mergers
Fig.~\ref{fig:params}  displays the  evolution  of  the periods  (left
panels)  and  eccentricities  (right  panels) of  the  binary  systems
studied here.  As  before, the top panels depict the  evolution of the
orbital  parameters for  $q_{\rm disk}=0.12$,  whilst the  bottom ones
show that  for the case in  which $q_{\rm disk}=0.10$ is  adopted.  As
can be clearly seen, in all  four simulations after each mass transfer
episode  the periods  of the  binary systems  decrease and  the orbits
become circularized.  These general  trends become more accentuated as
the evolution  proceeds.  Although  for $q_{\rm  disk}=0.10$ initially
the period decreases smoothly in an almost linearly way with a shallow
slope, after $t\sim  35 T_0$ the white dwarf almost  plunges on top of
the core of  the AGB star.  This  is in contrast with  what occurs for
$q_{\rm disk}=0.12$,  for which  the decrease  in the  orbital periods
occurs faster during the initial stages of the dynamical evolution, in
marked steps, and the white dwarf merges with the core of the AGB star
in very  few orbital periods, $\sim  2 T_0$. The right  panels of this
figure show that the eccentricity also decreases for increasing times.
For $q_{\rm disk}=0.12$ this occurs in marked steps, as it occurs with
the orbital periods,  a consequence of the  succesive passages through
the periastron,  while for  $q_{\rm disk}=0.10$ the  eccentricity also
decreases but oscillates  arroung a mean (decreasing)  value.  This is
caused by tidal forces.

\begin{table*}
\caption{Some  relevant characteristics  of the  merged remnants.   By
  columns: (1)  value of $q_{\rm disk}$  from equation \ref{eq:qdisk},
  (2) type  of run, hot (HC)  or cold (CC)  core of the AGB  star, (3)
  maximum temperature  attained during the dynamical  interaction, (4)
  nuclear energy released, (5) duration of the merger, (6) Mass of the
  merged remnant at the end of  the simulations, (7) mass of Keplerian
  disk at the  end of the simulations, (8) mas  of the extended shroud
  at the end of the simulations,  (9) mass of fall-back material, (10)
  ejected mass, (11)  maximum temperature of the  merged remnant, (12)
  maximaum  angular  speed  of  the  hot corona  at  the  end  of  the
  simulations.}
\label{tab:hydro}
\begin{center}
\begin{tabular}{cccccccccccc}
\hline
\hline
\noalign{\smallskip}
 $q_{\rm disk}$  &  Run  & $T_{\rm peak}$ & $E_{\rm nuc}$ & $\Delta t$ & $M_{\rm mr}$ & $M_{\rm disk}$ & $M_{\rm shroud}$  & $M_{\rm fb}$  & $M_{\rm ej}$  & $T_{\rm max}$ & $\omega_{\rm max}$  \\
   &       &       (K)      &     (erg)     &     (s)    & ($M_{\sun}$) & ($M_{\sun}$) & ($M_{\sun}$) & ($M_{\sun}$) & ($M_{\sun}$) &      (K)      &     (s$^{-1}$)        \\ 
\noalign{\smallskip}
\hline 
\hline
\noalign{\smallskip}
0.12 & HC & 9.23$\times 10^{8}$ & 1.13$\times 10^{39}$ & 18343 & 0.88 & 0.36 & 9.08$\times 10^{-2}$ & 2.19$\times 10^{-2}$ & 1.74$\times 10^{-2}$ & 3.56$\times 10^8$ & 0.23  \\
     & CC & 8.55$\times 10^{8}$ & 3.54$\times 10^{37}$ & 24784 & 0.88 & 0.35 & 9.84$\times 10^{-2}$ & 2.02$\times 10^{-2}$ & 1.80$\times 10^{-2}$ & 3.53$\times 10^8$ & 0.25  \\ 
\noalign{\smallskip}
\hline
0.10  & HC &  8.74$\times 10^8$ & 1.12$\times 10^{39}$ &  2599 & 0.93 & 0.39 & 4.86$\times 10^{-2}$ & 2.43$\times 10^{-3}$ & 7.37$\times 10^{-4}$ & 2.97$\times 10^8$ & 0.30  \\
      & CC &  8.18$\times 10^8$ & 4.77$\times 10^{37}$ &  2710 & 0.91 & 0.41 & 4.59$\times 10^{-2}$ & 2.56$\times 10^{-3}$ & 8.70$\times 10^{-4}$ & 3.10$\times 10^8$ & 0.32  \\
\noalign{\smallskip}
\noalign{\smallskip}
\hline
\hline
\end{tabular}
\end{center}
\end{table*}

Figure~\ref{fig:tem1} shows  the average  temperature of the  core for
the different  cases considered here.  To compute this  temperature we
averaged the temperature of those particles which were originally used
to model the core, and thus we  did not take into account the accreted
material from the disrupted white dwarf.  This is an important detail,
since the maximum  temperatures during the evolution  of the simulated
cases of coalescence are not reached in the core of the merged remnant
but in the hot corona formed  during the interaction --- see below. As
before, the left  panel and right panels  represent, respectively, the
cases  for  which  $q_{\rm  disk}=0.12$ and  $q_{\rm  disk}=0.10$  are
adopted. As can be seen, since for eccentric orbits the amount of mass
transferred during each dynamical  episode is larger, the temperatures
are  increased  more noticeably.   Actually,  the  temperature is  not
raised homogenously in the  entire core. Specifically, the temperature
of the outer layers of the  core is notably increased after every mass
transfer episode, while  the temperature of the very  deep interior of
the core remains nearly constant. As  it should be expected, the final
temperatures  of  the  hot  cores  are much  larger  than  their  cold
counterparts.

Table~\ref{tab:hydro}  shows  some  relevant  characteristics  of  the
stellar interactions studied  in this paper. In columns  three to five
of this table  we list, respectively, the  maximum temperature reached
during the  dynamical interaction,  $T_{\rm peak}$, the  total nuclear
energy released, $E_{\rm nuc}$, and  the total duration of the merger,
$\Delta t$.  The maximum temperature always  is reached at the base of
the accreted  layer.  The  nuclear energy  is essentially  released by
carbon  burning reactions  and  the $\alpha$  chains.   Note that  the
nuclear energy released in those mergers  in which the core of the AGB
is hot is typically 25 times larger than that released when cold cores
are involved.  The  reason for this is that for  mergers involving hot
AGB cores nuclear  reactions occur in a region  which is significantly
larger than  that in  which cold  cores are  considered.  Most  of the
nuclear reactions occur on top of the almost rigid surface of the core
of  the AGB  star, as  the material  of the  disrupted white  dwarf is
accreted, compressed and  heated.  A discussion of the  details of the
distribution of the chemical composition in  the remnant is out of the
scope  of this  paper, and  will be  presented elsewhere.  However, we
mention  here  that  these  results  are in  line  with  our  previous
simulations     of    merging     and    colliding     white    dwarfs
\citep{Loren-Aguilar2009,     Loren-Aguilar2010,     Aznar-Siguan2013,
Aznar-Siguan2014},  which  in  turn  agree with  the  calculations  of
\cite{Raskin2013} and \cite{Dan2014}.

Again, we emphasize  that the duration of the merger  not only depends
on the physical  characteristics of the coalescing stars,  but also on
the adopted  resolution.  The reason  for this  is that, as  it occurs
with DD mergers, the durations of the mass transfer episodes depend on
the  extent  to which  the  outer  layers of  the  donor  star can  be
resolved.  In most  practical situations the particle  masses are such
that  the  resolvable mass  transfer  rate  is already  a  substantial
fraction  of  the total  system  mass  ---  see \cite{Dan2011}  for  a
detailed discussion of  this issue.  Hence, these  durations should be
regarded as indicative, and can be  only used to compare the different
cases  studied  here,  which  were  computed  with  the  same  spatial
resolution and input physics.

Figs.~\ref{fig:time12},  \ref{fig:time10}  and  \ref{fig:macc}  reveal
that the  peak temperature is reached  at $t=1.8 T_0$ and  $t=2.4 T_0$
for the  hot and cold cores  of the AGB star  when $q_{\rm disk}=0.12$
and at  $t=38.1 T_0$ and  $t=40.3 T_0$ for  the case in  which $q_{\rm
disk}=0.10$.  When  this occurs  the fraction  of the  secondary white
dwarf that  has been accreted  onto the core of  the AGB star  is $\ga
20\%$ for $q_{\rm disk} = 0.12$,  whereas for $q_{\rm disk} = 0.10$ it
is $\la  10 \%$.  All  this has a  direct effect on  the thermonuclear
reaction rates, and consequently on  the total nuclear energy released
during the  interactions.  In  particular, while for  the case  of hot
cores the total nuclear energy  released is $\approx 10^{39}$~erg, for
the cold ones  it is much smaller, $\approx  10^{37}$~erg. The nuclear
energy  released  is   similar  for  the  eccentric   orbits  and  the
non-eccentric ones  for a fixed  temperature of the AGB  core, despite
the  difference   in  the   peak  temperatures  achieved   during  the
interaction.

Finally, we note that eccentric orbits ($q_{\rm disk}=0.12$) result in
mergers  with durations  of a  few hours,  much longer  than those  of
mergers with $q_{\rm disk}=0.10$, for  which the durations are smaller
than $\sim 1$ hour. However, we  remind that for the case of eccentric
orbits  the coalescence  occurs in  just a  couple of  initial orbital
periods, while for circular ones  the merger lasts for several initial
orbital periods.

\subsection{The remnants of the interaction}

\begin{figure*}
   \begin{center}
   {\includegraphics[width=0.47\textwidth, trim=0.1cm 0.1cm 0.1cm 0.1cm,clip=true]{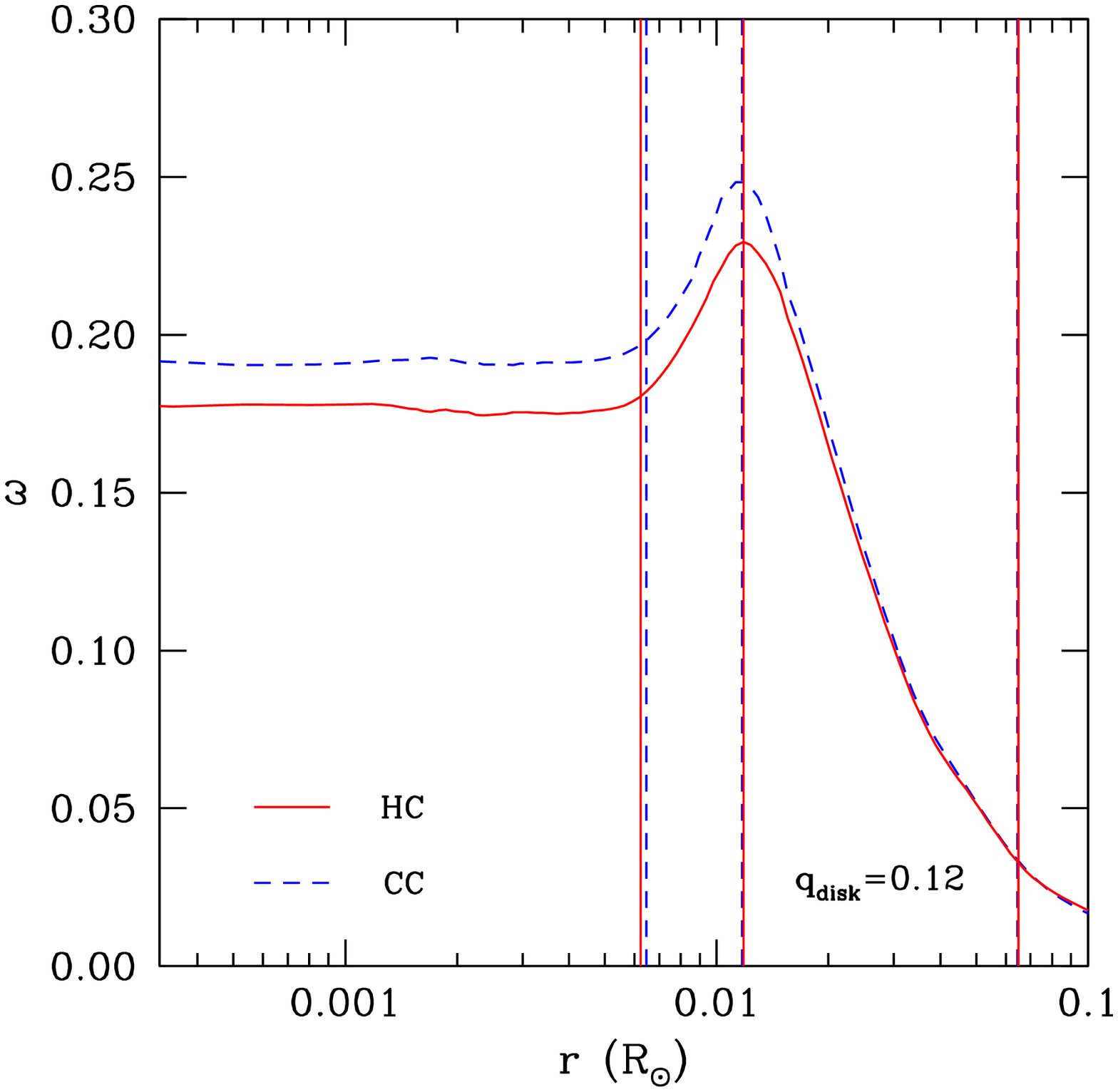}\hspace{0.3cm}
    \includegraphics[width=0.47\textwidth, trim=0.1cm 0.1cm 0.1cm 0.1cm,clip=true]{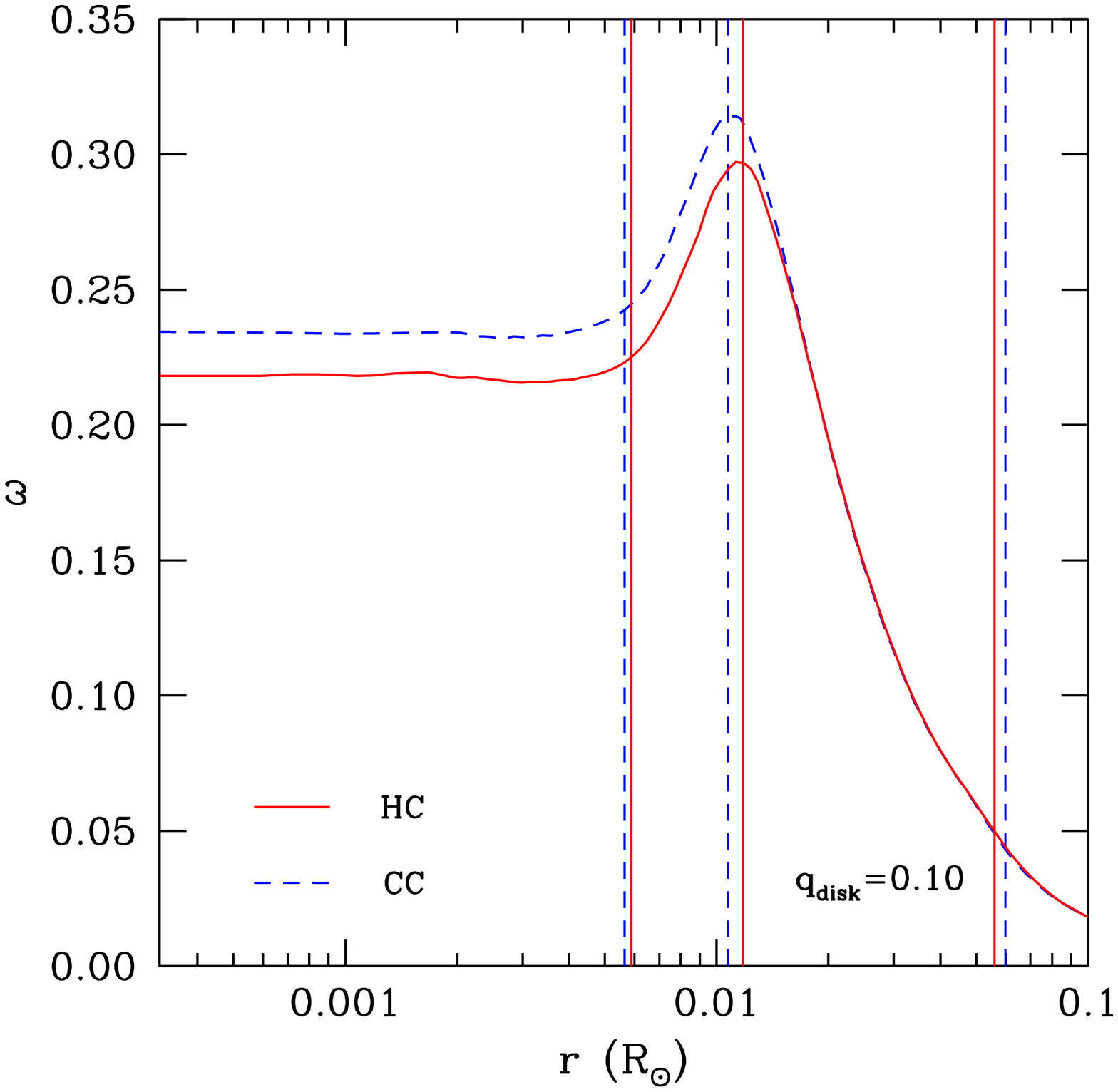}}\\ 
   \caption{Rotational  velocity   profiles  as  a  function   of  the
     spherical  radius in  solar  units, at  times $t/T_0=1.8115$  and
     2.4476 for the hot (solid red lines) and cold (dashed blue lines)
     AGB cores  of the  simulation in  which $q_{\rm  disk}=0.12$, and
     $t/T_0=40.2378$  and  41.9690  for  the  case  in  which  $q_{\rm
     disk}=0.10$ is adopted, respectively.   These times correspond to
     our last  computed models.   In the left  panel we  represent the
     profiles for $q_{\rm disk}=0.12$, while the right one corresponds
     to  $q_{\rm  disk}=0.10$.   Note  that for  large  distances  the
     spherical  radius is  not a  valid representation  of the  merged
     configuration, as  the assumption  of spherical symmetry  is only
     fulfilled to a good approximation in the innermost regions of the
     merged remnant. The leftmost solid  vertical lines show the outer
     boundaries  of the  respective degenerate  remnants.  The  middle
     vertical  lines show  the  location  of the  outer  edges of  the
     corresponding  coronae. The  rightmost  vertical  lines mark  the
     outer edges of the respective Keplerian disks.}
\label{fig:omega} 
   \end{center}
\end{figure*}

In this  section we  analyze with  the help  of Figs.~\ref{fig:omega},
\ref{fig:perf}, \ref{fig:cont} and Table~\ref{tab:hydro} the structure
of  the remnants  of the  interaction. The  general appearance  of the
remnants resulting from  the interaction is similar for  all the cases
studied  in this  paper,  and moreover  is similar  to  that found  in
previous simulations  of the DD  scenario for similar masses  --- see,
for instance,  \cite{Loren-Aguilar2009}, and references  therein.  The
remnants of the interactions when  simulations were stopped consist of
a central degenerate object which contains all the mass of the core of
the AGB  star and some accreted  mass. This central object  spins as a
rigid body and its angular momentum  arises from the conversion of the
orbital angular momentum of the binary system to rotational one.  This
core is  surrounded by  a corona, which  contains a  sizable fraction,
$\sim  18\%$ for  eccentric mergers  and $\approx  27\%$ for  circular
ones, of the  mass of the disrupted secondary (the  white dwarf).  The
material of the corona is hot --- a consequence of the material of the
disrupted  secondary  being compressed  on  top  of the  almost  rigid
surface of the core of the AGB star --- and rotates differentially ---
a consequence of  the several mass transfer episodes  occurring in the
dynamical interaction.  Most  of the material of  the secondary ($\sim
82\%$  and $\approx  73\%$, respectively,  for eccentric  and circular
mergers) that  is not accreted  on the central  object goes to  form a
thick accretion disk, with a Keplerian velocity profile. Additionally,
there is  an extended shroud made  of particles that have  orbits with
large inclinations with respect to  the orbital plane.  Also, some SPH
particles have highly  eccentric orbits and will  ultimately fall back
onto the  primary. Finally, we find  that very little mass  is ejected
from   the  system.    All  these   regions  are   clearly  shown   in
Fig.~\ref{fig:omega},  where  we  plot  the angular  velocity  of  the
remnant as a function of the  spherical radius for the remnants of the
interactions studied here. The vertical lines show from left to right,
respectively, the locations  of the outer edge of the  remnant, of the
hot, rapidly rotating corona and of the Keplerian disk.

Table~\ref{tab:hydro} lists  also some  of the characteristics  of the
merged remnant.   In column 6 we  list the mass of  the central merged
object  ($M_{\rm  mr}$) ---  that  is,  the sum  of  the  mass of  the
undisturbed AGB core and  of the hot corona --- while  in column 7 the
mass of  the Keplerian disk  ($M_{\rm disk}$) is listed.   The precise
location  of the  inner  edge of  the  hot corona  is  defined as  the
(spherical)  mass coordinate  for  which the  angular  velocity is  no
longer constant, while the outer boundary of this region is defined as
the point where  the profile of angular velocities of  the remnants is
the Keplerian.   In this table we  also show the mass  of the extended
shroud  ($M_{\rm shroud}$),  that  of the  fallback material  ($M_{\rm
fb}$), as well as the mass  ejected from the system ($M_{\rm ej}$) ---
columns  8, 9  and  10,  respectively.  As  mentioned,  the shroud  is
composed  of material  coming from  the disrupted  secondary that  has
approximately  a  spherical  distribution. The  fallback  material  is
composed of  particles belonging to  the debris region that  have very
eccentric orbits  as a result  of the extreme conditions  found during
the  very first  stages of  the dynamical  interaction, but  are still
gravitationally bound  to the  merged remnant.   Thus, it  is expected
that this material will ultimately  fall back onto the central remnant
at some later time \citep{Rosswog2007}.  To compute the amount of mass
in the  debris region that  will eventually  fall back on  the central
object    we    followed    the   same    procedure    described    in
\cite{Aznar-Siguan2014}.   Note that  there is  as well  some material
that is not gravitationally bound to the remnant and might escape, but
the merger  episodes are essentially  conservative.  We would  like to
remark that we do  not look for a possible formation  of jets and disk
winds  which,  in  the  DD  scenario influece  the  appearance  of  an
explosion    if     it    occurs    shortly    after     the    merger
\citep{Levanonetal2015}. Finally, for the sake of completeness we also
list the maximum  temperature ($T_{\rm max}$) and  the maximum angular
velocities ($\omega_{\rm max}$) of the  hot corona, columns 11 and 12,
respectively.

\subsubsection{The central compact object}

\begin{figure*}
   \begin{center}
   {\includegraphics[width=0.47\textwidth, trim=0.1cm 0.1cm 0.1cm 0.1cm,clip=true]{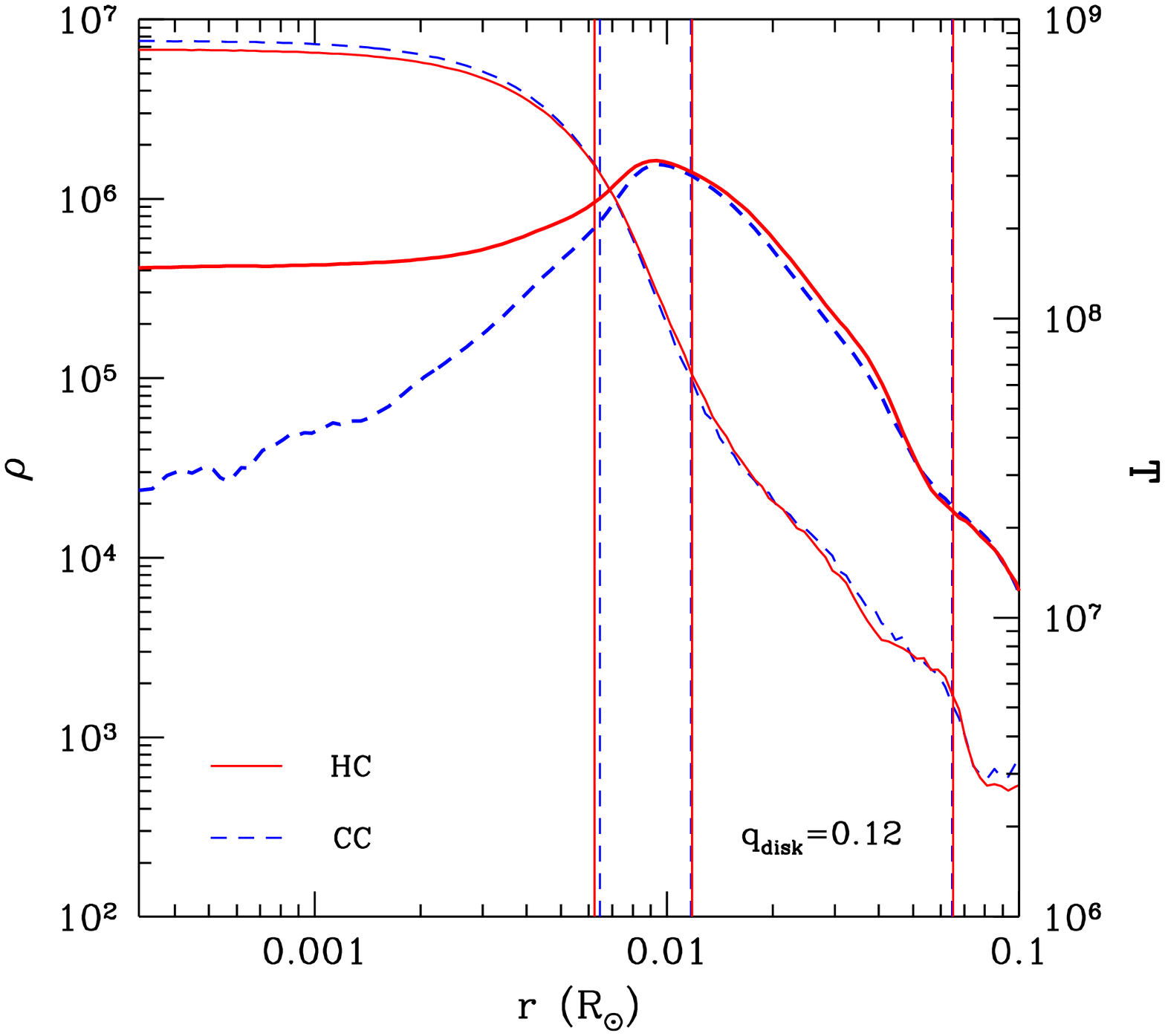}\hspace{0.3cm}
    \includegraphics[width=0.47\textwidth, trim=0.1cm 0.1cm 0.1cm 0.1cm,clip=true]{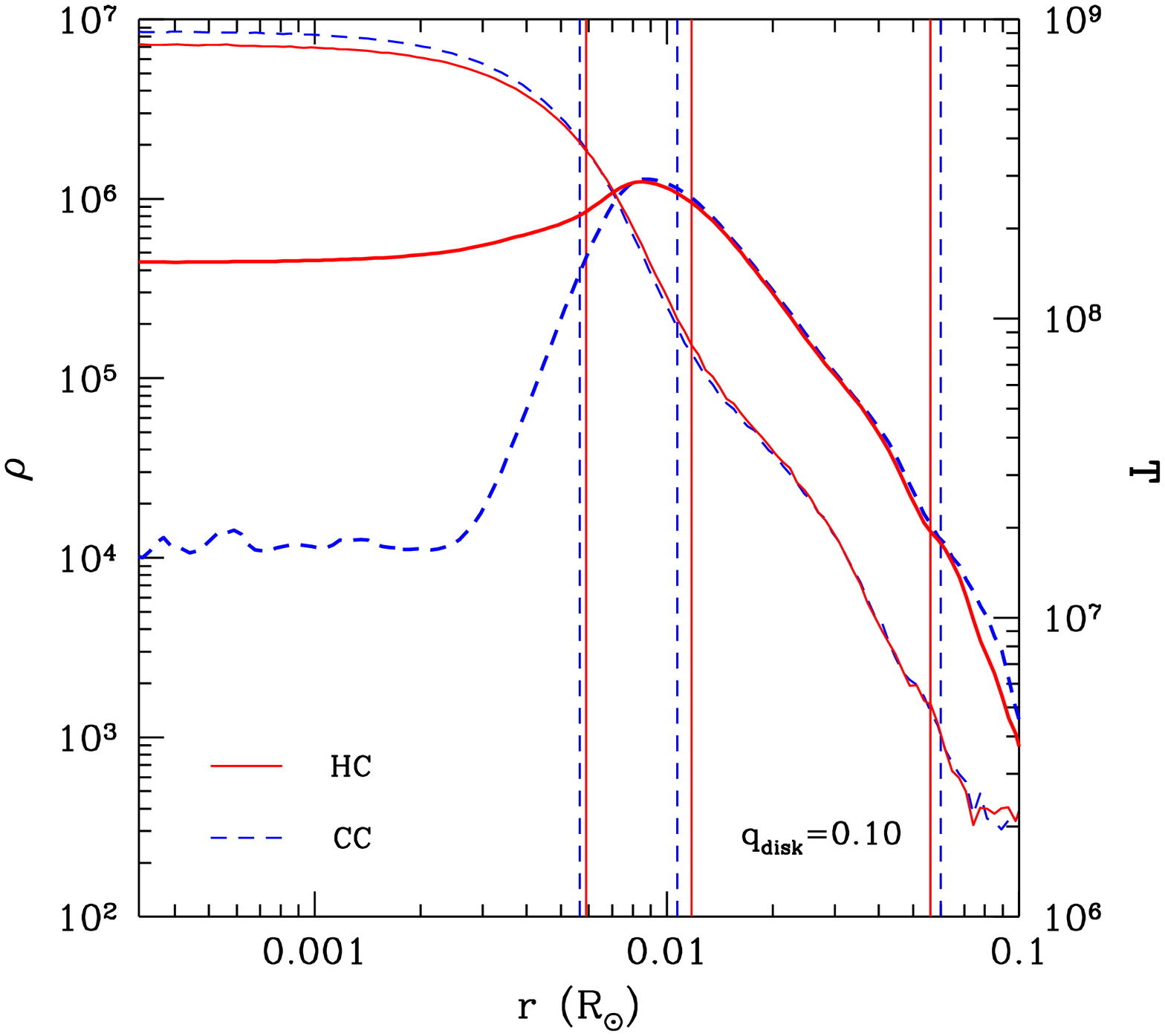}}\\ 
   \caption{Density  (thin   lines)  and  temperature   (thick  lines)
     profiles of the  central part of the remnants, at  the same times
     shown in Fig.~\ref{fig:omega}.  The left panel shows the profiles
     of the  hot (solid red  lines) and  cold (dashed blue  lines) AGB
     cores of the  simulation in which $q_{\rm  disk}=0.12$, while the
     right panel shows the same quantities for $q_{\rm disk}=0.10$. As
     before, the abscissa is the  spherical radius in solar units, and
     the  vertical   lines  show  the  same   locations  explained  in
     Fig.~\ref{fig:omega}.}
\label{fig:perf} 
   \end{center}
\end{figure*}

As can be seen in Table~\ref{tab:hydro}, the angular velocities of the
remnants do not  depend appreciably on the adopted  temperature of the
core  of the  AGB  star, although  there  are significant  differences
depending on the value of $q_{\rm  disk}$. The reason for this is that
in the coalescence studied here the  rigid rotation of the core of the
AGB star arises  from the conservation of angular  momentum.  That is,
the  orbital angular  momentum of  the pair  is basically  invested in
spinning up  the primary and  the previously described corona,  and to
form the  Keplerian disk.  Nevertheless,  the angular velocity  of the
remnants  of the  mergers in  which a  cold AGB  core is  involved are
somewhat larger  in both cases.  This  is due to the  slightly smaller
radii of  these configurations (see Fig.~\ref{fig:perf}  below), which
in turn is a consequence of their larger degeneracies.

We now compare the angular velocities obtained for different values of
$q_{\rm  disk}$.   As  shown in  Table~\ref{tab:hydro},  when  $q_{\rm
disk}=0.10$  is  adopted  the   angular  velocities  are  larger.   In
particular, we find that for this case the inner regions of the merged
remnant  rotate uniformly  at  a  speed $\omega\simeq  0.22$~s$^{-1}$,
whereas  in the  corona  a maximum  angular  velocity of  $\omega_{\rm
max}\simeq 0.31$~s$^{-1}$ is  reached.  For the case  in which $q_{\rm
disk}=0.12$ these angular  velocities are, respectively, $\omega\simeq
0.18$~s$^{-1}$, and  $\omega_{\rm max}\simeq  0.24$~s$^{-1}$, somewhat
smaller  than  those of  the  case  in  which $q_{\rm  disk}=0.10$  is
employed. The orbital  angular momentum of the  binary systems studied
here is $J  = 5.96 \times 10^{50}$~erg~s for  $q_{\rm disk}=0.12$, and
$J = 4.63  \times 10^{50}$~erg~s when $q_{\rm  disk}=0.10$.  Since the
merger  is almost  conservative (very  few particles  acquire energies
large enough to escape from  the merged remnant) this angular momentum
is distributed essentially in spinning up the core of the AGB star, in
the differential rotation of the hot corona and in Keplerian disk.

For  a  hot  merger  their  respective  angular  momenta  are  $J_{\rm
core}\simeq 1.4\times 10^{49}$~erg~s, $J_{\rm corona} \simeq 4.4\times
10^{49}$~erg~s  and $J_{\rm  disk}\simeq 2.8\times  10^{50}$~erg~s for
$q_{\rm disk}=0.12$, and $J_{\rm core}\simeq 1.5\times 10^{49}$~erg~s,
$J_{\rm   corona}   \simeq   6.7\times  10^{49}$~erg~s   and   $J_{\rm
disk}\simeq   3.2\times   10^{50}$~erg~s  for   $q_{\rm   disk}=0.10$,
respectively. For the case of a cold merger these values do not differ
much. Hence, the  orbital angular momentum of the  eccentric merger is
larger than that of the circular one, and moreover the angular momenta
are distributed in  a different way in both  cases.  Specifically, the
angular momentum of  the corona for $q_{\rm  disk}=0.12$ is noticeably
smaller  than that  of  the  merger in  which  $q_{\rm disk}=0.10$  is
adopted, but the  mass of their respective coronae do  not differ that
much.   Also the  angular momentum  of the  newly formed  disk of  the
eccentric merger  is significantly smaller  than that of  the circular
one. Consequently,  more angular momentum  is stored in  these (outer)
regions  for the  case in  which $q_{\rm  disk}=0.10$ is  adopted when
compared to the  case in which $q_{\rm disk}=0.12$.   Finally, for the
the simulations in  which $q_{\rm disk}=0.12$ is  employed the ejected
mass is  considerably larger than  that of  the case in  which $q_{\rm
disk}=0.10$ is  used.  This mass  carries some angular  momentum.  All
this  explains why  the  merged remnant  spins at  a  slower rate  for
$q_{\rm  disk}=0.12$.  All  this  is a  consequence  of the  dynamical
evolution during the  merger.  We remind that  for $q_{\rm disk}=0.12$
the evolution proceeds in abrupt  steps, while for $q_{\rm disk}=0.10$
mass is accreted in a more gentle way.

Figure~\ref{fig:perf} shows the density (dashed lines) and temperature
(solid  lines) profiles,  for the  case in  which $q_{\rm  disk}=0.12$
(left panel) and the for the  case in which $q_{\rm disk}=0.10$ (right
panel).  The maximum  temperature ($T_{\rm max}$) in  all the remnants
is very similar, close to  $3\times 10^8$~K, being slightly higher for
the remnants of the eccentric  mergers.  This can be easily understood
in  terms  of  the  evolution  during the  last  merger  episode.   In
particular, it has  been already mentioned that  for eccentric mergers
the mass  transfer episodes  are more  dramatic than  for the  case of
nearly circular orbits.  Hence, the  matter of the disrupted secondary
white  dwarf  acquires  larger  accelerations  and,  consequently,  is
compressed more violently  on the surface of the  primary star.  Thus,
the temperatures  of the external  layers of the  merged configuration
are larger, and a hotter corona is obtained.

\begin{figure*}
   \begin{center}
      {\includegraphics[width=0.47\textwidth, trim=0.1cm 0.1cm 0.1cm 0.1cm,clip=true]{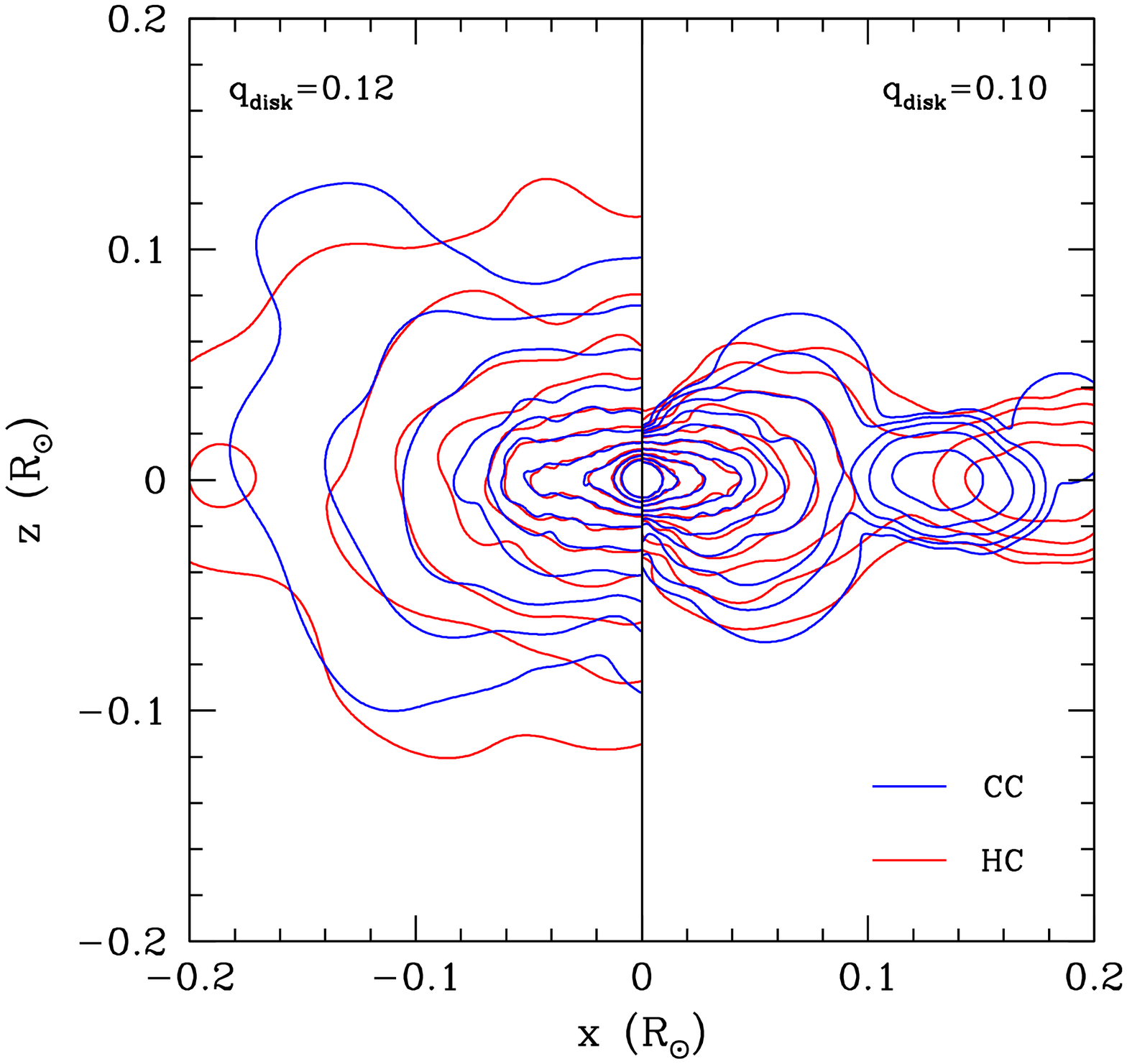}\hspace{0.3cm}
       \includegraphics[width=0.47\textwidth, trim=0.1cm 0.1cm 0.1cm 0.1cm,clip=true]{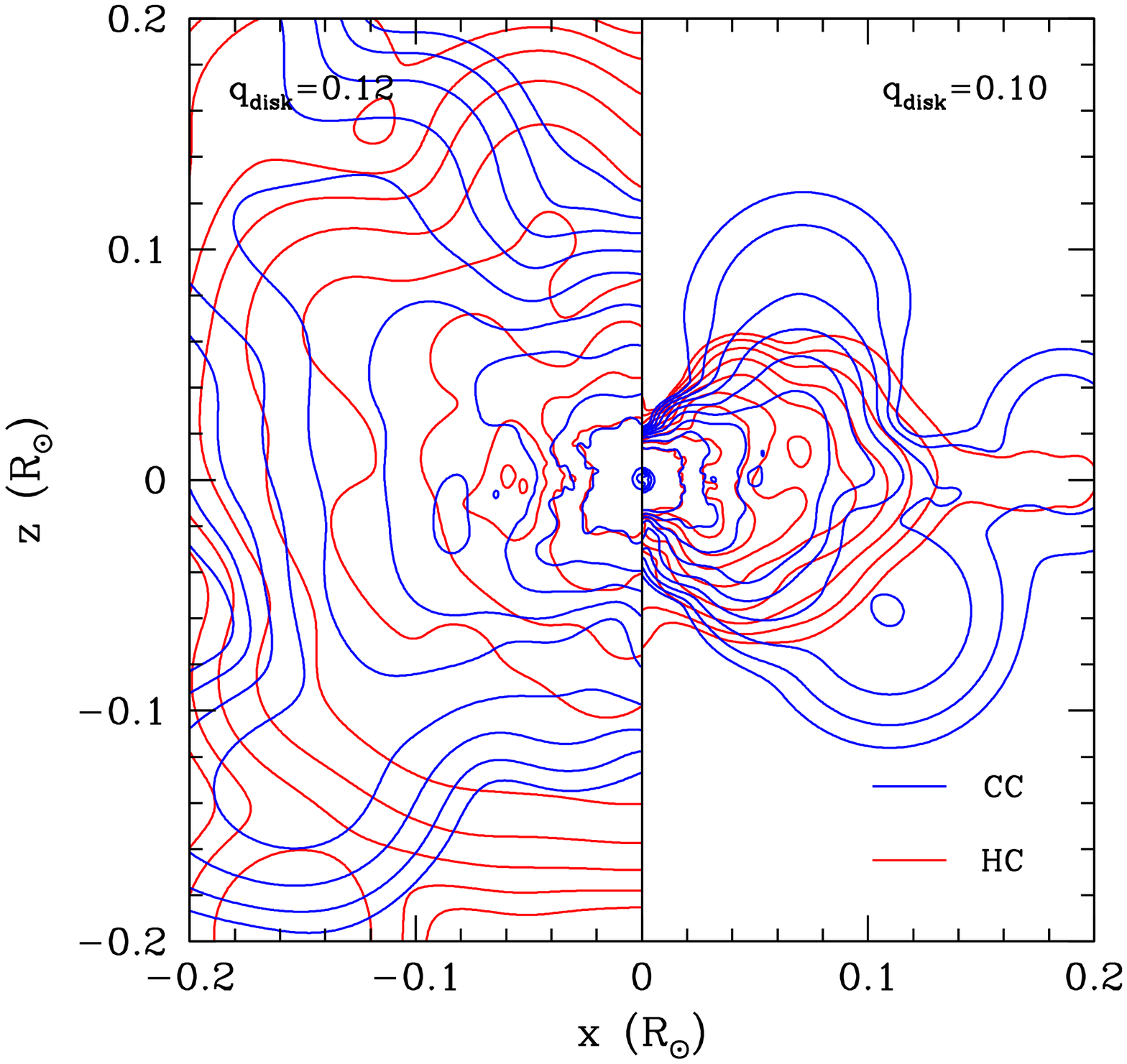}}\\
   \caption{Left panel:  density contour lines of  the remnants across
     the  meridional plane.   Right panel:  temperature contour  lines
     across  the same  section.   These density  and temperature  maps
     correspond  to our  last  computed models,  and the  evolutionary
     times  are   the  same  of  Fig.~\ref{fig:perf}.    The  remnants
     represented with negative $x$  coordinates correspond to the case
     in which  $q_{\rm disk}=0.12$, while  for positive values  of $x$
     the case in  which $q_{\rm disk}=0.10$ is  represented.  For both
     cases the outermost contour of the density profile corresponds to
     a density  of 1~g~cm$^{-3}$ and the  successive contours increase
     logarithmically inwards in constant steps of $\log\rho=0.62$. The
     contour  of  maximum  density   corresponds  to  $\sim  3.8\times
     10^5$~g~cm$^{-3}$.   For the  temperature  the outermost  contour
     corresponds to $\sim 3 \times 10^5$~K and the successive contours
     increase inwards logarithmically in  steps of $\log T=0.33$.  The
     maximum  temperature  contour  corresponds to  $\sim  1.4  \times
     10^8$~K.  This  temperature is reached  in the hot  corona, while
     the innermost regions of the merged remnant in the case of a core
     with   small    temperature   merger    are   colder    ---   see
     Fig.~\ref{fig:tem1}. See the online edition  of the journal for a
     color version of this figure.}
\label{fig:cont}  
   \end{center}
\end{figure*}

The  stronger interaction  of  the accreted  matter  with the  primary
during  the  successive  mass  transfer  episodes  leaves  also  clear
imprints  on the  structure  of  the internal  regions  of the  merged
remnant. Specifically,  for the case  of eccentric mergers in  which a
cold  core of  the  AGB star  is considered,  the  temperature of  the
outermost regions of the initially  isothermal core begins to increase
at earlier  times during  the coalescence, when  compared to  those in
which a circular  orbit is involved.  Consequently, by the  end of the
merger process  the core of this  remnant is no longer  isothermal. On
the contrary, the core in the  case of a circular orbit remains nearly
isothermal  ---   see  Fig.~\ref{fig:perf}.  Finally,  owing   to  the
substantial degeneracy  of the material of  the core of the  AGB star,
the density  profiles obtained in  all the four  simulations presented
here are very similar.

\subsubsection{The debris region}

It has  been already discussed  that the  temperature of the  AGB core
barely plays a role in the  dynamical evolution of the merger process,
and in determining  the structure of the merged  remnant. Instead, the
most important parameter is the  eccentricity of the initial orbits of
both components  of the binary  system. This is  true as well  for the
debris region  formed around the  merged remnants.  The masses  of the
fallback and  of the ejected  material are, respectively,  $M_{\rm fb}
\simeq 2.1\times 10^{-2}\, M_{\sun}$ and $M_{\rm ej} \simeq 1.8 \times
10^{-2}\, M_{\sun}$  for eccentric mergers, while  for circular orbits
the masses  are much smaller,  $M_{\rm fb} \simeq  2.5\times 10^{-3}\,
M_{\sun}$ and  $M_{\rm ej}  \simeq 8.1\times 10^{-4}\,  M_{\sun}$. The
larger fallback  and ejected masses  in the case of  eccentric mergers
are a  consequence of the  longer durations (or,  equivalently, larger
orbital   separations)  of   the  mass   transfer  episodes   ---  see
Figs.~\ref{fig:macc} and \ref{fig:params}.

As  well,  the  masses  of   the  respective  shrouds  are  relatively
different, $M_{\rm  shroud} \simeq  9.2\times 10^{-2}\,  M_{\sun}$ for
$q_{\rm  disk}=0.12$ and  $M_{\rm shroud}  \simeq 4.7\times  10^{-2}\,
M_{\sun}$ for $q_{\rm disk}=0.10$.  This is again a consequence of the
very  different number  of mass  transfer episodes  occurring in  both
simulations.  Actually, for $q_{\rm disk}=0.12$ we find that some mass
of the  disrupted white dwarf flows  around the region located  in the
opposite direction  of the line  connecting both stars of  the system,
and acquires  high velocities  when colliding  with the  primary star,
forming this extended  shroud.  The size of this  extended region does
not depend  much on the  temperature adopted for  the core of  the AGB
star. This can be seen in Fig.~\ref{fig:cont}, where the density (left
panel) and temperature (right panel) contours of the merged remnant in
the meridional plane are displayed.  In this figure the sub-panel with
negative values of the $x$-coordinate  corresponds to the case $q_{\rm
disk}=0.12$, whereas that with positive values shows the same contours
for the case in which $q_{\rm disk}=0.10$ is adopted.

As  can be  seen  in  Fig.~\ref{fig:cont}, both  the  density and  the
temperature decrease  as the  distance to  the central  compact object
increases, independently if  we consider the remnants  of eccentric or
circular mergers. However the decline rate of both the density and the
temperature for eccentric mergers is smaller than for circular orbits,
resulting in a more extended  debris region, which moreover is hotter.
Additionally,  the  density  and  temperature  profiles  also  have  a
different behavior near the polar regions of the central remnant.  For
mergers  arising from  circular  orbits  there is  a  region near  the
$z$-axis (the vertical axis in this  figure) where the density and the
temperature contour lines converge towards the poles, in contrast with
what happens for the case of  eccentric mergers, for which this region
is absent, and the contour lines are perpendicular to this axis. These
differences  between  both  profiles  are a  consequence  of  how  the
material of the disrupted secondary  white dwarf is distributed on top
of  the primary  (the  core  of the  AGB  star)  during the  dynamical
interaction.  

The final infall of the disrupted secondary onto the unaltered primary
for   $q_{\rm   disk}=0.10$   seen   in   Figs.~\ref{fig:time10}   and
\ref{fig:orb},  does  not allow  matter  to  be redistributed  on  the
primary, but  instead concentrates it  in the orbital  plane. Besides,
the circular density contours that appear  at the right of the remnant
correspond to a spiral arm of  matter infalling to the primary.  Since
SPH codes are not able  to follow properly angular momentum transport,
we stopped  the simulations  at a reasonable  time after  the complete
disruption of  the secondary.  In particular,  for $q_{\rm disk}=0.12$
we stopped  the simulations  at $t/T_0=1.8115$ and  $t/T_0=2.4476$ for
the hot and cold merger,  respectively.  For $q_{\rm disk}=0.10$ these
times are, respectively, $t/T_0=40.2378$  and $t/T_0=41.9690$. In this
case, the  secondary had no  time to  be totally distributed  over the
primary and the spiral arm formed  during the infalling phase is still
present. The  more flat  structure for  $q_{\rm disk}=0.10$,  which is
more  similar to  that  usually found  in the  simulations  of the  DD
scenario,  might lead  to jets  and disk  winds that  might alter  the
observations    if    explosion    occurs   shortly    after    merger
\citep{Levanonetal2015}.  Finally, the  outer edges  of the  Keplerian
disks formed during  the merger process are located  where the density
distribution in the equatorial plane has no longer axial symmetry.  In
all cases this occurs at $\sim 0.06 \,R_{\sun}$.

\section{Summary and conclusions}
\label{sec:conclusion}

In this paper we have simulated  the coalescence of a white dwarf with
the  core of  a massive  AGB  star, in  the context  of the  so-called
core-degenerate scenario  for SNe  Ia. Specifically, we  have computed
the  merger  of  an  otherwise  typical white  dwarf  of  mass  $0.6\,
M_{\sun}$ with  the core  of an  AGB star  of mass  $0.77\, M_{\sun}$,
which is a typical case \citep{Kashi2011}.  This has been done for two
temperatures of  the AGB  core, $T=10^6$~K and  $10^8$~K, and  for two
assumptions  about the  mass  of  the disk  formed  during the  common
envelope phase.  In particular, we have  studied a first case in which
the  mass of  the disk  is  $M_{\rm disk}=  0.12 (M_{\rm  core}+M_{\rm
WD})$, which corresponds to the critical mass ratio for which a merger
is guaranteed as a consequence of the interaction with the disk formed
during  the  common  envelope  phase. The  second  case  studied  here
corresponds to  a binary  system surviving  the common  envelope phase
with a mass ratio 0.10, which is below the critical one, and for which
the  merger is  driven by  the emission  of gravitational  waves.  The
first of these cases results in  a pair with a highly eccentric orbit,
whereas for  the second one  the orbits of  the members of  the binary
system are circular, and similar to those expected for the DD scenario
for SNe Ia.

Our SPH simulations show that temperature  of the core of the AGB star
has little influence on the overall dynamical evolution of the merging
process  and on  the characteristics  of the  merger remnant,  but the
initial eccentricity of  the binary system has a large  impact on both
the dynamical  evolution of the  merger and  on the properties  of the
merged object. In particular, in both cases the white dwarf is totally
disrupted  in a  series  of  mass transfer  episodes,  which occur  at
closest approach.  However, the duration of the coalescence is notably
different in both  cases. Specifically, we found that for  the case of
an eccentric merger the disruption of the secondary white dwarf occurs
in very  few orbital periods, whereas  for the case of  the mergers in
which circular orbits are adopted  this process lasts for many orbital
periods.  Nevertheless, the initial orbital  periods in both cases are
very different  --- 10,126~s and  65~s, respectively --- so  the total
time required  to disrupt the  secondary white  dwarf turns out  to be
larger for the case of eccentric orbits.

In all the four cases studied in this paper the coalescence process is
almost conservative, and little mass is ejected from the system, $\sim
2\times 10^{-2}\, M_{\sun}$ at  most.  Nevertheless, eccentric mergers
eject $\sim  25$ times more mass  than that ejected by  binary systems
with  circular  orbits.  The  peak  temperatures  achieved during  the
merging process are rather high, of the order of $8\times 10^8$~K, but
not enough to power a prompt detonation upon merger.  As for the final
temperature of  the merger remnant  it is interesting to  realize that
for   the   cases  studied   here   the   temperature  profile   peaks
off-center. The  possibility that  carbon is  ignited in  the remnant,
thus producing  a rapidly spinning  oxygen-neon white dwarf,  has been
discussed  by \cite{SN2004},  \cite{YL2004},  and \cite{YL2005}.   Our
results do  not support a  prompt off-center ignition, since  the peak
temperature obtained  at the end  of our  simulations is too  low, and
moreover the  temperature peaks  in a region  where the  densities are
relatively small.   However, even though  carbon in our models  is not
ignited immediately during  the merger process, it may  occur at later
times, when the high entropy material  begins to cool and compress the
material near the temperature peak.  Nevertheless, this depends on the
not yet well known viscous  timescale of the disk, which characterizes
the transport of disk mass  inwards and angular momentum outwards, and
on the thermal timescale of the merger product.  Thus, the post-merger
evolution    has     received    recently     considerable    interest
\citep{VanKerkwijk2010, Schwab2012, Shen2012, GB2012, Ji2013, Zhu2013,
Beloborodov14}  but a  clear consensus  on whether  a powerful  carbon
ignition can develop at later times has not emerged yet.
  
The  structure of  the merged  remnant consists  of a  central compact
object containing all  the mass of the  core of the AGB  star and some
mass from the disrupted secondary star,  which is surrounded by a hot,
differentially rotating corona.  This corona  is made of about 18\% of
the  mass of  the disrupted  white dwarf,  for the  case of  eccentric
mergers,  and  somewhat  larger   ($\sim  27\%$)  for  circular  ones.
Finally, surrounding  this central  compact object  there is  a heavy,
rapidly-rotating Keplerian  disk, and  an extended shroud  of material
which are formed by the rest of  the mass of the disrupted white dwarf
that  has  approximately  ellipsoidal shape.   This  configuration  is
essentially the same found for  DD mergers.  However, the structure of
the debris region  is different depending on  the initial eccentricity
of the merger.  For mergers  with initially circular orbits the matter
of the disrupted white dwarf  concentrates in regions near the orbital
plane, and  little mass  is found  in the  polar regions,  whereas for
eccentric mergers  part of the  mass of  the disrupted white  dwarf is
distributed  along   the  central  compact  object,   and  the  merged
configuration adopts an ellipsoidal shape.  In the debris region there
is also some  fraction of SPH particles with  highly eccentric orbits.
This material  will ultimately interact  with matter in  the Keplerian
disk and will fall back onto the central compact remnant.  The mass of
this fallback material is typically or the order of $2\times 10^{-2}\,
M_{\sun}$ for the case of the binary systems with eccentric orbit, and
ten times  smaller for those with  circular ones.  In all  cases, this
central  object  rotates  as  a   rigid  body,  with  typical  angular
velocities $\sim 0.18$~s$^{-1}$ for the  case of eccentric mergers and
$\sim  0.22$~s$^{-1}$ for  circular ones.   These rotation  velocities
arise from  the conversion of  orbital angular momentum  to rotational
one, as it happens for DD mergers.

We would  like to emphasize that  if the disk, fallback  material, and
the material  of the extended  shroud are  finally accreted, as  it is
commonly thought,  the mass  of the remnant  would be  $\approx 1.35\,
M_{\sun}$.   If this  is the  case, after  cooling and  losing angular
momentum this remnant might  eventually explode.  Of course, assessing
such possibility requires  modelling the viscous phase  of the merger,
which transports disk mass inwards and angular momentum outwards. This
phase has typical timescales ranging from $10^3$~s to $10^4$~s, during
which the  disk might  launch jets and  winds \citep{Levanonetal2015},
and it cannot be followed  using SPH techniques.  If, nonetheless, all
this mass is accreted and the remnant explodes the delay would be very
long, up  to millions of  years, or even longer  \citep{Ilkov2012}. In
particular, for accretion  of helium onto a  carbon-oxygen white dwarf
in a  stable regime,  in a recent  study \cite{Piersanti2014}  found a
delay  of  up  to  millions  of  years,  in  agreement  with  previous
calculations  of   this  kind  \citep{Nomoto1985,   SN2004,  Yoon2007,
ShenBildsten2009}. It  remains to be  checked if the same  might occur
for the CD merger process.

In summary, this  first set of simulations of the  CD scenario for SNe
Ia demonstrates  the need of exploring  in more detail the  effects of
the  eccentricity (and  possibly of  the temperature)  on the  overall
dynamical evolution of the merger and  on the properties of the merged
configurations obtained in this scenario, especially when more massive
stars  are  involved.  The  simulations  presented  here are  a  first
important step in this direction.   Future works will undoubtely allow
us to  characterize and  place tight constraints  on the  most salient
features of  this scenario, and  may possibly help in  quantifying the
fraction of these  mergers that contributes to the total  SNe Ia rate,
thus filling a gap and complementing  what we already learned from the
existing simulations of the DD scenario.

\section*{Acknowledgements}

We thank an anonymous referee for comments that improved and clarified
the presentation of our results.  This research was supported by MCINN
grant AYA2011--23102, and by the European Union FEDER funds.

\bibliographystyle{mn2e}
\bibliography{noam}

\end{document}